\documentclass[twocolumn]{aastex631}

\usepackage{subfiles}
\usepackage{amsmath,amssymb}
\usepackage{gensymb}
\usepackage{rotating}
\usepackage[hyphens]{url}
\usepackage{graphicx}
\usepackage{wrapfig}
\usepackage{color}

\newcommand{\spitzer}{{\it{Spitzer}}}
\newcommand{\sofia}{{\it{SOFIA}}}

\mathchardef\mhyphen="2D 
\shorttitle{FORCAST Galactic Center Catalog}
\shortauthors{Cotera \& Hankins, et al.}

\graphicspath{{./}{figures/}}

\begin{document}


\title{SOFIA/FORCAST Galactic Center Source Catalog}

\correspondingauthor{Angela S. Cotera}
\email{ascotera@gmail.com, acotera@seti.org}

\author[0000-0002-0786-7307]{Angela S. Cotera}
\affiliation{SETI Institute, 189 Bernardo Ave., Mountain View, CA 94043, USA}

\author[0000-0001-9315-8437]{Matthew J. Hankins}
\affil{Arkansas Tech University, 215 West O Street, Russellville, Arkansas 72801 USA}

\collaboration{13}{(SOFIA Galactic Center Legacy Project)}


\author{John Bally}
\affil{Department of Astrophysical and Planetary Sciences, University of Colorado, 389 UCB, Boulder, CO 80309, USA}

\author[0000-0003-0410-4504 ]{Ashley T. Barnes}
\affiliation{European Southern Observatory (ESO), Karl-Schwarzschild-Straße 2, 85748 Garching, Germany}

\author{Cara D. Battersby}
\affiliation{University of Connecticut, Department of Physics, 196A Auditorium Road, Unit 3046, Storrs, CT 06269}

\author{H Perry Hatchfield}
\affiliation{Jet Propulsion Laboratory, California Institute of Technology, 4800 Oak Grove Drive, Pasadena, CA, 91109, USA}

\author[0000-0002-3856-8385]{Terry L. Herter}
\affil{Department of Astronomy, Cornell University, Space Sciences Bldg, Ithaca, NY 14853-6801, USA}

\author[0000-0003-0778-0321]{Ryan M.\ Lau}
\affil{NSF’s NOIRLab, 950 N. Cherry Ave., Tucson, AZ 85719, USA}

\author{Steven N. Longmore}
\affil{Astrophysics Research Institute, Liverpool John Moores University, 146 Brownlow Hill, Liverpool L3 5RF, UK}

\author[0000-0001-8782-1992]{Elisabeth A. C. Mills}
\affil{Department of Physics and Astronomy, University of Kansas, 1251 Wescoe Hall Dr., Lawrence, KS 66045, USA}

\author[0000-0002-6753-2066]{Mark R. Morris}
\affil{Dept. of Physics and Astronomy, University of California, Los Angeles, CA 90095-1547, USA}

\author{James T. Radomski}
\affil{SOFIA-USRA, NASA Ames Research Center, MS 232-12, Moffett Field, CA 94035, USA}

\author[0000-0001-8095-4610]{Janet P. Simpson}
\affiliation{SETI Institute, 189 Bernardo Ave., Mountain View, CA 94043, USA}

\author{Zachary Stephens}
\affil{Arkansas Tech University, 215 West O Street, Russellville, Arkansas 72801 USA}

\author{Daniel L. Walker}
\affil{UK ALMA Regional Centre Node, Jodrell Bank Centre for Astrophysics, The University of Manchester, Manchester M13 9PL, UK}



\begin{abstract}
The central regions of the Milky Way constitute a unique laboratory for a wide swath of astrophysical studies, consequently the inner $\sim$400 pc has been the target of numerous large surveys at all accessible wavelengths. 
In this paper we present a catalog of sources at 25 and 37 \micron\ located within all of the regions observed with the \sofia/FORCAST instrument in the inner $\sim$200 pc of the Galaxy.  The majority of the observations were obtained as part of the \sofia\ Cycle 7 Galactic Center Legacy program survey, which was designed to complement the \spitzer/MIPS 24 \micron\ catalog in regions saturated in the MIPS observations. Due to the wide variety of source types captured by our observations at 25 and 37 \micron, we do not limit the FORCAST source catalog to unresolved point sources, or treat all sources as if they are point-like sources. The catalog includes all detectable sources in the regions, resulting in a catalog of 950 sources, including point sources, compact sources, and extended sources. We also provide the user with metrics to discriminate between the source types.  
\end{abstract}


\section{Introduction} \label{intro} The center of the Milky Way is galactically unique with physical properties more like those typically found in starburst galaxies than in the Galactic disk \citep[e.g.][]{Kruijssen2013}. In the region known as the Central Molecular Zone (CMZ) the molecular gas densities are high \citep[$\sim 10^4$ cm$^{-3}$; e.g.,][]{Mills2018}, gas and dust temperatures are highly variable \citep[$\sim~20-400$~K, e.g.,][]{Ginsburg2016,Barnes2017,Tang2021}, and large turbulent mach numbers are derived \citep[$\sim30$; e.g.,][]{Kruijssen2014}. This region is also host to three of the most massive stellar clusters in the galaxy - the Arches \citep{Cotera1996}, the Quintuplet \citep{Okuda1990,Nagata1990}, and the central cluster \citep{Krabbe1991,Krabbe1995}. These clusters, along with numerous isolated massive stars spread throughout the region, are the source of the radiation and winds which ionize and shape the surrounding interstellar medium on both small and large scales. Finally, no summary of the unique nature of the Galactic Center (GC) is complete without noting the impact of the supermassive black hole Sgr A$^*$, which contributes significantly to the energetics of the region \citep[e.g.][]{Zhao2016,Morris2023}. Past activity of Sgr A$^*$ may be responsible for large-scale features such as the x-ray `chimneys' discovered by the \textit{XMM-Newton} satellite \citep{Ponti2019,Ponti2021}. Due to the complexity and unique nature of the GC, there is a wealth of observational and theoretical studies of the region as discussed in detail in \citet{Bryant2021} and \citet{Henshaw2023}, and references therein.

 Our relative proximity to the GC \citep[8.18$\pm$0.01 kpc;][]{Gravity2019}, enables us to obtain data with higher resolution and sensitivity compared to what is possible in any external galaxy. However, because observations of the GC are made through the Galactic plane, optical extinction toward the region \citep[A$_V\sim30$; e.g.][]{Fritz2011} is so significant that optical surveys of the region are not possible. Consequently, our understanding of the stellar population depends primarily upon observations at infrared wavelengths. In particular, as observational capabilities advanced over the past two decades, large surveys of the inner $\sim$500 pc have been completed, with most currently publicly available \citep[see Table 1,][]{Bryant2021}.  The increases in coverage, resolution, and sensitivities of these surveys have revolutionized our understanding of this enigmatic region.  

There have been a handful of large mid-infrared ($\sim$4--40 \micron) surveys which have produced searchable catalogs.  In particular the \spitzer/MIPS point source catalog assembled by \citep{hinz2009} provided photometry for the MIPS 24 \micron\ survey of the inner 1.5$\degree\times$8.0$\degree$. The high sensitivity of \spitzer/MIPS at 24 \micron, however, resulted in significant portions of the survey being unusable due to saturation \citep[see Figure 1,][]{Hankins2020}. More recently, \citet{Hankins2020} presented observations of the inner $\sim200$ pc of the GC, using \sofia/FORCAST at 25 and 37 \micron\ as part of the \sofia\ Galactic Center Survey Legacy program.  Those data were combined with previously available \sofia/FORCAST 25 and 37 \micron\ observations to create a large mosaic, which was discussed in \citet{Hankins2020}. Since the publication of that paper, additional fields have been observed with FORCAST in the same configuration as the Legacy Survey and are presented below.  

In this paper we use the best available observations of the Galactic Center obtained with \sofia/FORCAST at 25 and 37 \micron\ to construct a comprehensive source catalog for these observations. The structure of this paper is as follows: in section 2 we discuss the observations that are included in the catalog.  In section 3 we discuss the methods we use to derive the measurements provided in our source catalog, including fluxes, uncertainties, quality flags, and extinction estimates. In section 4 we provide analysis of our findings, including our completeness limits, astrometric uncertainties, comparisons with the \citet{hinz2009} \spitzer/MIPS data for overlapping sources, and a preliminary discussion of the additional parameter space that can be investigated using the catalog data.  We summarize the development of the catalog in section 5.  Detailed information on the contents of the full online catalog and software developed to produce the catalog, both of which are publicly available, are provided in Appendices \ref{howto} and \ref{code} respectively.

\begin{sidewaysfigure*}
\vspace*{3.5in}
\centering
\includegraphics[scale=0.35, ]{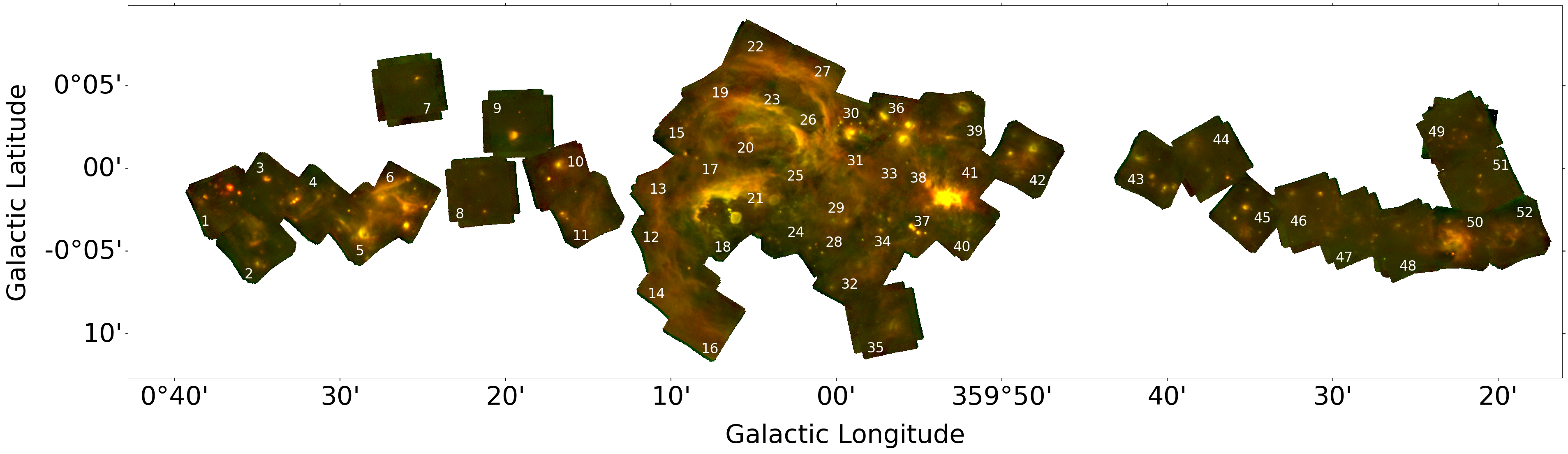}
\includegraphics[scale=0.35, ]{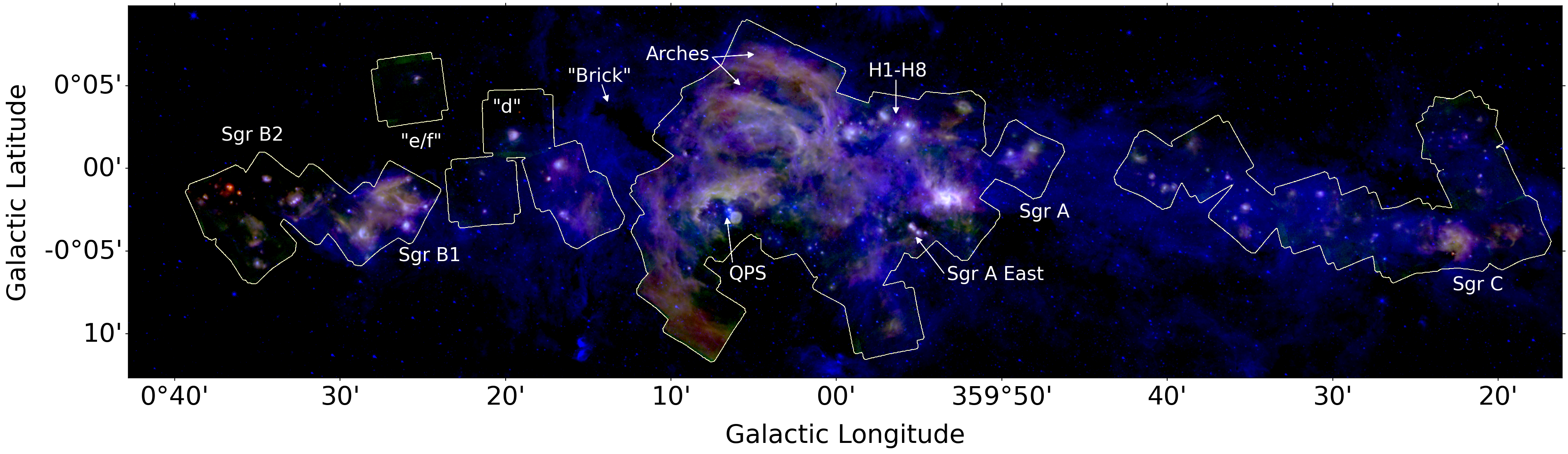}
\caption{{\it{Upper:}} Mosaic of all currently available \sofia/FORCAST images in the Galactic Center at 25 and 37 \micron. The  source catalog presented here is derived from observations of the individual fields.  The numbers correspond to the SFGC ({\it SOFIA}/FORCAST Galactic Center) Survey and Catalog fields as listed in Table \ref{obs_tab}.  The 37 \micron\ mosaic image is red, the 25 \micron\ image is green.  {\it{Lower:}} Same as upper image, with the addition of the {\it{Spitzer}}/IRAC 8 \micron\ image of the Galatic Center in blue.  Well known regions of interest are labeled.  \label{fig1}}
\end{sidewaysfigure*}

\begin{deluxetable*} {ccl|cc|cc|cl} 
\tabletypesize{\footnotesize}
\tablecaption{Observations\label{obs_tab}}
\tablehead{
\colhead{SFGC}&\colhead{Obs}&\colhead{}&\multicolumn{2}{c}{Coordinates}&\multicolumn{2}{c}{Int. Time (s)} & \colhead{}&\colhead{} \\  
\colhead{Field} & \colhead{Date} & \colhead{Fits Header Object Name} & \colhead{l} & \colhead{b} & \colhead{25 \micron } & \colhead {37 \micron} & \colhead{AOR ID} &\colhead{Other Names} 
}
\startdata
1&1-Jul-19&Galactic Center Field\_11&0.674&-0.051&210.2&176.1&07\_0189\_11&Sgr B1\\
2&1-Jul-19&Galactic Center Field\_10&0.641&-0.092&527.9&480.1&07\_0189\_10&Sgr B1\\
3&1-Jul-19&Galactic Center Field\_9&0.631&-0.041&614.4&625.7&07\_0189\_9&Sgr B1\\
4&4-Jul-19&Galactic Center Field\_8&0.584&-0.053&432.9&404.0&07\_0189\_8&Sgr B1\\
5&4-Jul-19&Galactic Center Field\_12&0.534&-0.072&414.2&386.6&07\_0189\_35&Sgr B2\\
6&1-Jul-19&Galactic Center Field\_13&0.496&-0.051&522.2&531.9&07\_0189\_36&Sgr B2\\
{\bf{7}}&{\bf{2-Jul-21}}&{\bf{Field\_A-1}}&{\bf{0.487}}&{\bf{0.063}}&{\bf{261.1}}&{\bf{241.4}}&{\bf{09\_0216\_1}}& \\
{\bf{8}}&{\bf{2-Jul-21}}&{\bf{Field\_B-1}}&{\bf{0.413}}&{\bf{-0.040}}&{\bf{556.4}}&{\bf{514.2}}&{\bf{09\_0216\_2}}& \\
{\bf{9}}&{\bf{2-Jul-21}}&{\bf{Field\_C-1}}&{\bf{0.377}}&{\bf{0.028}}&{\bf{567.7}}&{\bf{524.7}}&{\bf{09\_0216\_3}}& \\
10&9-Jul-19&Galactic Center Field\_7&0.337&-0.024&217.1&191.7&07\_0189\_7& \\
11&9-Jul-19&Galactic Center Field\_6&0.310&-0.058&458.1&415.4&07\_0189\_6& \\
12&4-Jul-19&Galactic Center Field\_U&0.223&-0.089&468.2&437.0&07\_0189\_32& \\
13&2-Jul-19&Galactic Center Field\_ &0.222&-0.044&491.9&479.5&07\_0189\_31& \\
14&9-Jul-19&Galactic Center Field\_V&0.216&-0.135&504.9&457.8&07\_0189\_33& \\
15&2-Jul-19&Galactic Center Field\_R&0.198&0.022&426.8&386.5&07\_0189\_29& \\
16&9-Jul-19&Galactic Center Field\_W&0.190&-0.169&482.4&500.3&07\_0189\_34& \\
17&8-Jul-19&Galactic Center Field\_S&0.182&-0.018&469.3&430.2&07\_0189\_30& \\
18&10-Jul-19&Galactic Center Field\_Y&0.174&-0.068&504.3&470.6&07\_0189\_38&QPS\\
19&4-Jul-15&Arches NE Cyc3&0.173&0.053&166.0&856.2&70\_0300\_23& \\
20&3-Jul-19&Galactic Center Field\_Q&0.145&0.006&518.2&500.3&07\_0189\_28&Arches Cluster\\
21&10-Jul-19&Galactic Center Field\_X&0.141&-0.045&483.7&451.4&07\_0189\_37&Sickle\\
22&3-Jul-15&Arches NW Cyc3&0.137&0.094&284.5&246.3&70\_0300\_20& \\
23&4-Jul-15&Arches E Cyc3&0.130&0.039&213.4&173.2&70\_0300\_26&Filament W1\\
24&2-Jul-19&Galactic Center Field\_P&0.102&-0.073&524.7&511.5&07\_0189\_27& \\
25&3-Jul-19&Galactic Center Field\_O&0.101&-0.023&472.4&450.8&07\_0189\_26& \\
26&7-Jul-15&Arches SE Cyc3&0.093&0.020&224.5&210.5&70\_0300\_29&Filament W1\\
27&4-Jul-15&Arches W Cyc3&0.085&0.070&253.0&248.0&70\_0300\_17&Filament W2\\
28&9-Jul-19&Galactic Center Field\_M&0.065&-0.089&493.4&447.3&07\_0189\_24& \\
29&2-Jul-19&Galactic Center Field\_L&0.055&-0.053&523.8&450.7&07\_0189\_23& \\
30&13-Jun-15&Region H North&0.054&0.027&302.5&345.2&70\_0300\_15&Filament W2\\
31&3-Jul-19&Galactic Center Field\_K&0.051&-0.008&504.9&396.5&07\_0189\_22& \\
32&8-Jul-19&Galactic Center Field\_I&0.039&-0.124&504.9&457.8&07\_0189\_20& \\
33&3-Jul-19&Galactic Center Field\_G&0.015&-0.022&491.5&399.4&07\_0189\_18& \\
34&2-Jul-19&Galactic Center Field\_H&0.012&-0.079&526.8&490.5&07\_0189\_19& \\
{\bf{35}}&{\bf{1-Jul-21}}&{\bf{Field\_F}}&{\bf{0.009}}&{\bf{-0.169}}&{\bf{372}}&{\bf{392.7}}&{\bf{09\_0216\_6}}& \\
36&13-Jun-15&Region H South&359.996&0.026&92.5&75.2&70\_0300\_12&H1-H8\\
37&8-Jul-19&Galactic Center Field\_F&359.975&-0.064&472.0&432.7&07\_0189\_17&Sgr A\\
38&3-Jul-19&Galactic Center Field\_E&359.970&-0.019&393.2&319.5&07\_0189\_16&Sgr A\\
39&4-Jul-19&Galactic Center Field\_B&359.942&0.027&486.3&453.8&07\_0189\_13& \\
40&8-Jul-19&Galactic Center Field\_D&359.934&-0.068&458.1&415.4&07\_0189\_15&Sgr A\\
41&1-Jul-19&Galactic Center Field\_C&359.930&-0.019&439.5&368.1&07\_0189\_14&Sgr A\\
42&1-Jul-19&Galactic Center Field\_A&359.867&-0.007&477.8&400.1&07\_0189\_12& \\
43&11-Jul-19&Galactic Center Field\_5&359.737&-0.018&461.4&359.6&07\_0189\_5& \\
{\bf{44}}&{\bf{7-Jul-21}}&{\bf{Field\_I}}&{\bf{359.681}}&{\bf{-0.007}}&{\bf{539.7}}&{\bf{539.7}}&{\bf{09\_0216\_9}}& \\
45&11-Jul-19&Galactic Center Field\_3&359.641&-0.062&403.8&314.6&07\_0189\_3& \\
{\bf{46}}&{\bf{26-May-22}}&{\bf{Field\_J}}&{\bf{359.58}}&{\bf{-0.061}}&{\bf{544.5}}&{\bf{544.5}}&{\bf{09\_0216\_10}}& \\
{\bf{47}}&{\bf{26-May-22}}&{\bf{Field\_K}}&{\bf{359.538}}&{\bf{-0.076}}&{\bf{513.6}}&{\bf{513.6}}&{\bf{09\_0216\_11}}& \\
{\bf{48}}&{\bf{20-May-22}}&{\bf{Field\_L}}&{\bf{359.483}}&{\bf{-0.089}}&{\bf{487.1}}&{\bf{487.1}}&{\bf{09\_0216\_12}}& \\
{\bf{49}}&{\bf{7-Jul-21}}&{\bf{Field\_N-1}}&{\bf{359.429}}&{\bf{0.019}}&{\bf{575.4}}&{\bf{575.4}}&{\bf{09\_0216\_14}}& \\
50&8-Jul-19&Galactic Center Field\_2&359.429&-0.087&384.6&384.6&07\_0189\_2&Sgr C\\
{\bf{51}}&{\bf{25-May-22}}&{\bf{Field\_O}}&{\bf{359.408}}&{\bf{-0.032}}&{\bf{354.2}}&{\bf{354.2}}&{\bf{09\_0216\_15}}& \\
52&8-Jul-19&Galactic Center Field\_1&359.376&-0.080&402.1&402.1&07\_0189\_1&Sgr C  
\enddata
\end{deluxetable*}

\begin{figure*}
\centering
\includegraphics[scale=0.15]{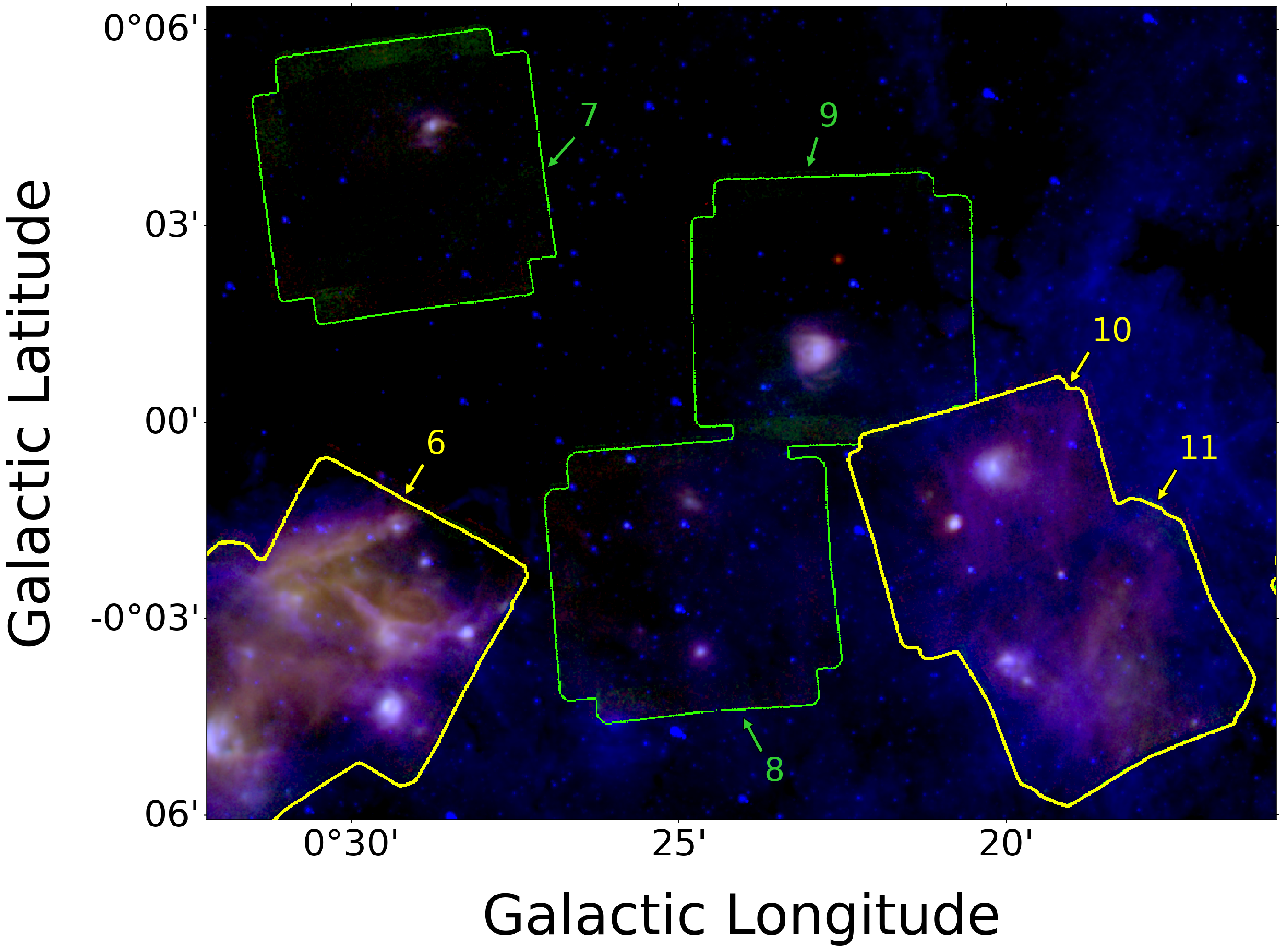}
\includegraphics[scale=0.16]{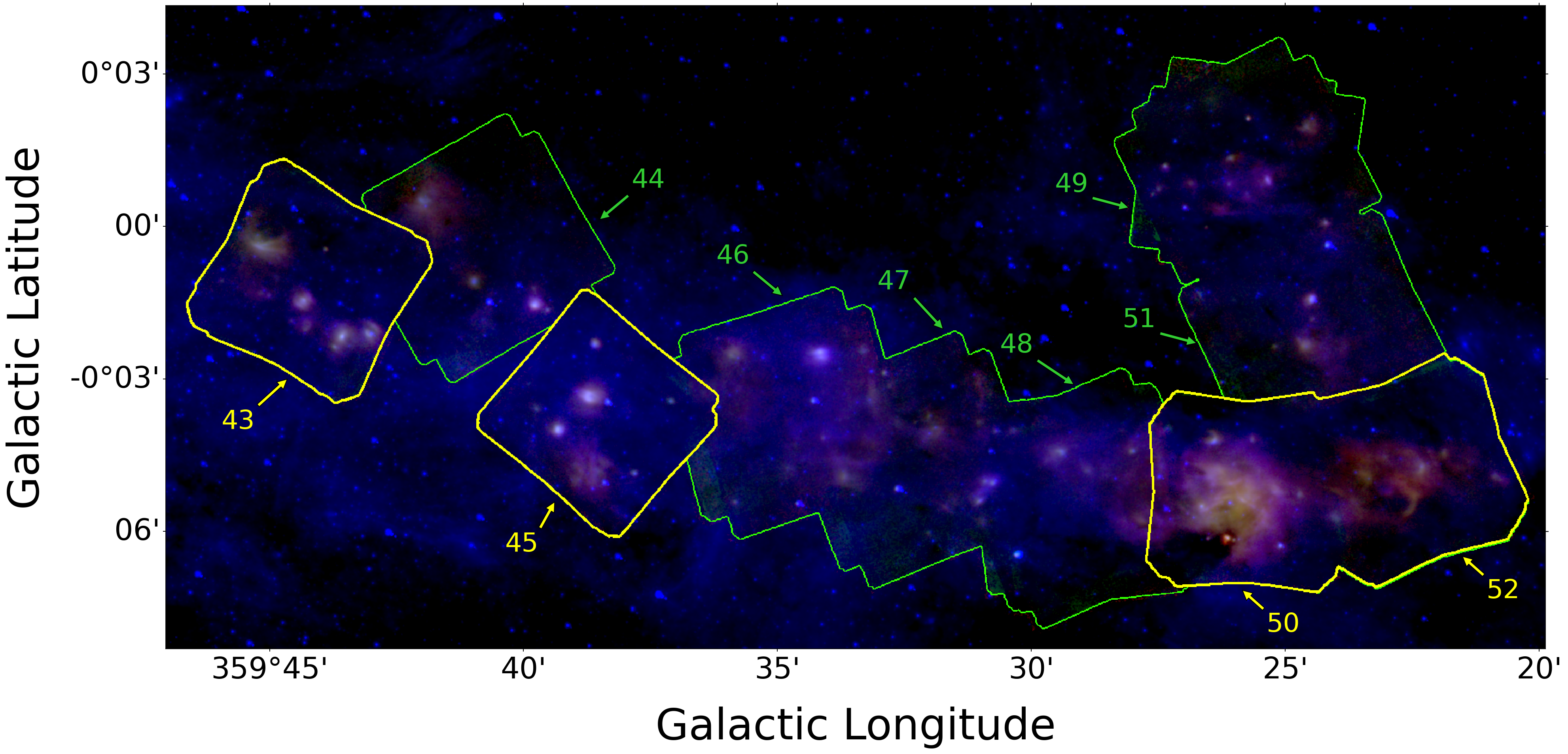}
\caption{{Additional regions observed in Cycle 9.  Regions observed as part of Cycle 7 are outlined in yellow, regions observed in Cycle 9 are outlined in green.  All regions are labeled with their corresponding field number in Table \ref{obs_tab}.  Red is the 37 \micron\ image, green is 25 \micron, and blue is the \spitzer/IRAC 8 \micron\ image.   \it{Left:}} The regions at positive longitudes between Sgr B and Sgr A. Sgr B1 is seen in the lower left.  {\it{Right:}} The regions at negative longitudes between Sgr A and Sgr C. Sgr C is seen in the lower right.  \label{cycle9}}
\end{figure*}

\vspace{-18pt}
\section{Observations and Data Reduction} \label{obs}
Observations were obtained on the 2.5 m telescope aboard the Stratospheric Observatory for Infrared Astronomy \citep[\sofia;][]{Young2012} using the FORCAST instrument \citep{Herter2012}. FORCAST is a $256 \times 256$ pixel dual-channel, wide-field mid-infrared camera with a field of view (FOV) of $3.4'\,\times\,3.2'$ and a plate scale of $0.768''$ per pixel. The two channels consist of a short-wavelength camera (SWC) operating at $5-25$ \micron\ and a long-wavelength camera (LWC) operating at 28 -- 40 $\mu\mathrm{m}$. The two channels can be observed simultaneously using the dichroic beam-splitter, the configuration employed for all of the observations presented in this paper. The nominal point spread function (PSF) for an unresolved source has a full width at half maximum (FWHM) of $\sim2.3\arcsec$ at 25 \micron\ and $\sim3.4 \arcsec$ at 37 \micron; however, this performance was not uniformly achieved due to observational anomalies as discussed in \cite{Hankins2020}, and inherent small FORCAST PSF variations from flight to flight.   

The source catalog presented here is derived from the best quality FORCAST observations taken of the Galactic Center at 25 and 37 \micron.  The catalog covers all of the regions observed with FORCAST at these wavelengths, and includes all observations except those obtained as part of the SOFIA early science program. Several of the early science program fields were observed again in Cycle 7 to obtain higher quality data and to ensure consistency with the full Legacy Survey.  The observations we use were obtained as part of three \sofia/FORCAST programs: 70-300 (PI: Herter), 07-189 (PI: Hankins), and 09-216 (PI: Hankins). The first observations were obtained in June 2015, and the last were completed in May 2022.  A total of 52 fields were observed, and are summarized in Table \ref{obs_tab}. The majority of the fields (35) were obtained as part the SOFIA Galactic Center Survey Legacy program, with an additional 7 fields obtained prior to the survey program.  All of these observations were combined into a single map of the GC and are discussed in greater detail in the Legacy program overview paper \citep{Hankins2020}. An additional 10 fields were obtained as part of Cycle 9, and are presented here for the first time with further details provided in section \ref{cycle9obs}.

Table \ref{obs_tab} uses the image fits header information to provide the observation date, the fits file object name, central coordinates, and integration times. We have used boldface to highlight observations taken in Cycle 9.  Since we are combining data from three separate programs, with diverse naming conventions (as reflected in the fits file object names), here we superimpose a single naming convention based on Galactic location, beginning with Field 1 located at Sgr B2, going east to Field 52 just past Sgr C.  A mosaic of all the images and the corresponding field numbers is presented in Figure \ref{fig1}.  

As can be seen in Table \ref{obs_tab}, the observations have a significant range of integration times. The legacy survey data and Cycle 9 data were designed to obtain consistent results, but small alterations were necessary due to flight constraints. For the majority of the fields the observations were designed to achieve a 
nominal 5$\sigma$ point source depth of 250 mJy at 25 \micron, which is equivalent to a 3$\sigma$ extended source depth of 1200 MJy sr$^{-1}$. Based on the integration times determined by the 25 \micron\ observations, the simultaneously obtained 37
\micron\ 5$\sigma$ point source depth was 550 mJy. The 25 \micron\ imaging depth is comfortably below the MIPS hard saturation limit for the existing 24 \micron\ GC map ($\sim$400 mJy for point sources or $\sim$2300 MJy sr$^{-1}$ for extended emission). 

\subsection{Cycle 9 observations}\label{cycle9obs}
In Cycle 9, we observed an additional 10 fields: 3 are between Sgr B and Sgr A, 5 are between Sgr A and Sgr C, and 2 are north of Sgr C as shown in Figures \ref{fig1} and \ref{cycle9}. Three of these fields were part of the original plan for the  Cycle 7 legacy program, but were not observed at that time due to scheduling constraints. The goal of the Cycle 9 program was to expand data coverage with \sofia/FORCAST to ensure a more comprehensive map of the GC at 25 and 37 \micron, particularly in regions with compact sources identified in other observations that are detectable with FORCAST but did not have robust 24--25 \micron\ data (e.g., saturated objects in the \spitzer/MIPS observations) or had prominent \textit{Herschel} 70 $\mu$m sources.  
To give an example, there are numerous less-well-studied mid-IR sources at negative Galactic longitudes between Sgr A and Sgr C which were observed with the cycle 9 program. The asymmetry of sources in the GC has long been known, and several of these fields were selected to help reduce possible observational bias in the study of star formation throughout the region.  

\subsection{Data Reduction}
The data reduction steps are presented in detail in \citet{Hankins2020}, here we provide a brief summary. All observations were processed using the pipeline steps described in \cite{Herter2013}. Images from each individual pointing were combined using the \textit{SOFIA} Data Pipeline software REDUX \citep{Clarke2015} to construct the preliminary FORCAST Level 4 image mosaics presented here. Both the Level 3 and 4 data products from this program are available for download via 
the NASA/IPAC Infrared Science Archive (IRSA).\footnote{\url{https://irsa.ipac.caltech.edu/frontpage/}}

Creating the mosaics for this data set involved several challenges which are discussed in detail in \citet{Hankins2020}.  For the catalog presented here, it is important to note that background matching was performed between adjacent fields in the mosaics, and this information was used in processing the final images for individual fields. As part of data reduction steps for C2NC2 imaging with \sofia/FORCAST, it is customary to force the background near the edges of the images to be approximately zero.\footnote{Any sufficiently large-scale emission features compared to the instrument FOV are subtracted away as part of the nodding and chopping process intrinsic to the observations} The complex emission features commonly encountered in the GC (e.g., molecular clouds), mean that the usual step of forcing the observations to zero median background can result in a mismatch between the flux levels in neighboring fields. Thus, we have carefully examined all overlapping regions between fields and appropriately scaled background levels to ensure agreement between the data in the large mosaic. There is still the potential for an overall DC offset in the imaging data for this program; however, this does not adversely impact the derived source parameters presented here since any flux associated with large scale features is removed as part of the background subtraction performed on all fields (see section \ref{detection}). 

We also note that telescope pointing for fields observed with FORCAST, and the resultant astrometric measurements, are only accurate to within a few pixels ($\sim$1\arcsec).  There are also well known issues with distortion corrections with the instrument \citep{Adams2012}, especially in C2NC2 mode, which relies on use of the entire instrument field-of-view. Therefore, astrometry was absolutely calibrated using the available \textit{Spitzer} and \textit{MSX} data by matching up the centroids of point sources in common between those maps and the \sofia\ data. The astrometric solutions for the observations, along with any adjustments related to the discussion above, were used as part of the catalog input. In Section \ref{astrometry}, we explore the astrometric accuracy of the catalog sources in greater depth.  

\begin{figure}
\includegraphics[scale=0.44]{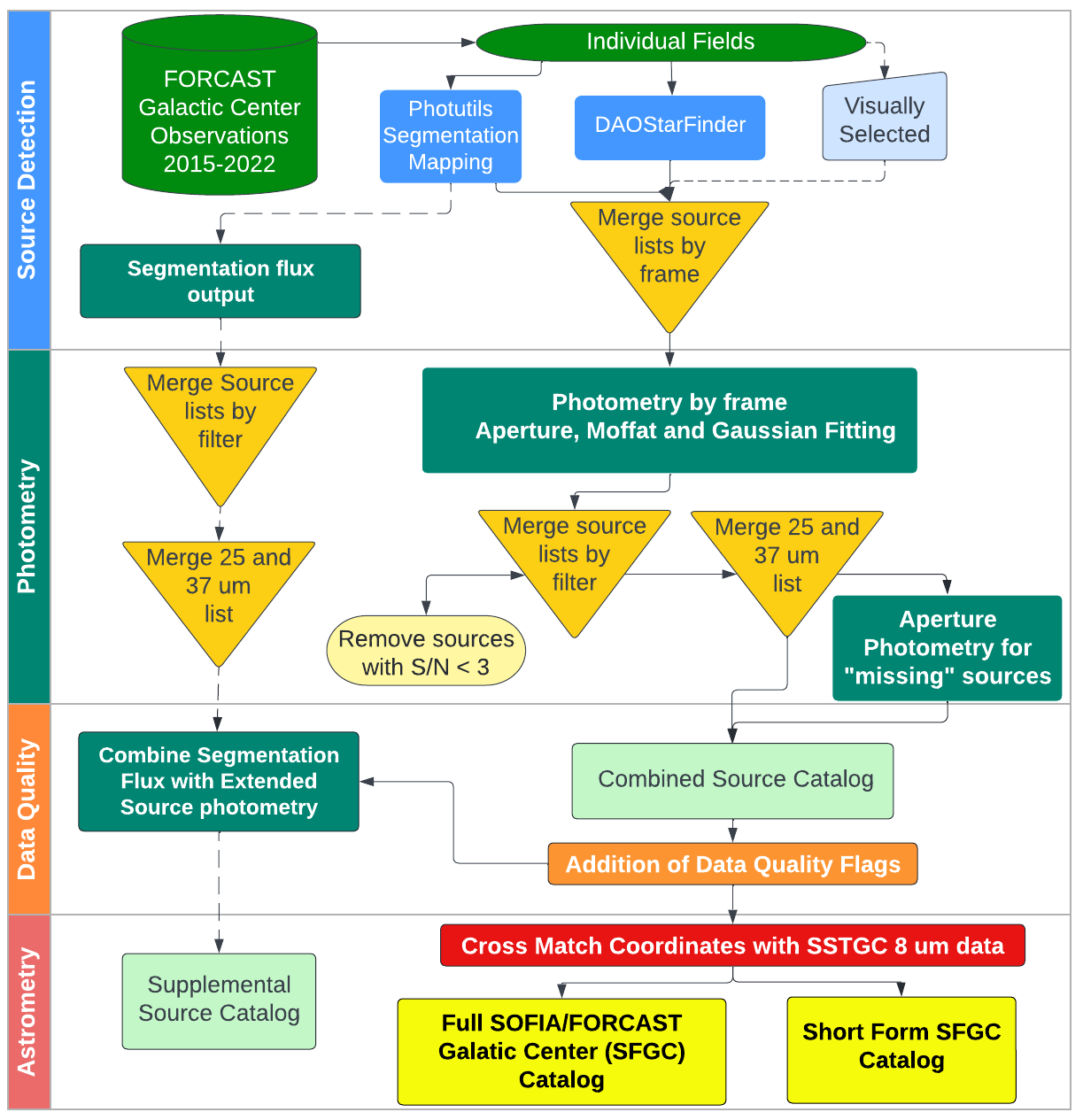}
\caption{Flowchart illustrating the methodology used to create the SOFIA/FORCAST Galactic Center (SFGC) catalogs. \label{flowchart}}
\end{figure}

\section{Creation of the FORCAST Source Catalog} \label{data}

Figure \ref{flowchart} presents the steps taken to create the source catalog presented in this paper, including additional plans for a supplemental catalog of very extended sources which is beyond the scope of the current paper. As shown in the figure, processing steps for identifying and extracting source fluxes are performed on individual frames. This choice was motivated by the fact that the observations of individual fields vary in terms of data quality \citep[e.g., see][]{Hankins2020} and also integration times. Furthermore, the \sofia/FORCAST PSF is not stable, and is known to vary from flight to flight. These issues present many challenges which preclude working with the mosaic images (e.g. Figure \ref{fig1}) to produce the source catalog for this program.

We note that in the Galactic Center, the interrelationship of objects ranging from stellar point-like sources to highly extended nebular objects is scientifically important, particularly when seeking to create a comprehensive picture of this highly complex region. Consequently, we decided early on to include all source types in our FORCAST catalog rather than focusing strictly on point-like sources. This is of particular importance given that the beam size of \sofia/FORCAST at the GC is $\sim$0.1 pc, such that numerous compact \ion{H}{2} regions are resolved throughout the region \citep[e.g.,][]{Hankins2019}.

Details on the processing steps are explained more fully in the following sections, but here we provide a brief synopsis. The process starts by running multiple detection algorithms on each individual field and then combining the results to produce source lists for each field. Next, multiple photometric measurements are performed based on the derived source coordinates, including aperture photometry and model-fitting routines which use 2D Moffat and 2D Gaussian functions. We determined that a comprehensive catalog necessitates inclusion of multiple, varied source measurements because of the wide variety of source types present in the observations. There is a continuum of source types in this region ranging from point-like, to marginally extended, to highly extended, and one extraction method was incapable of capturing robust fluxes for this diversity of source types.  

After source detection and photometry, quality cuts are used to remove any low SNR sources. The outputs for individual fields are then merged to create two source catalogs at 25 \micron\ and 37 \micron. From here the 25 and 37 \micron\ catalogs are cross-matched and combined. Photometry is performed again for any sources that have only one wavelength measurement at this stage to ensure that there are measurements at both wavelengths for all sources or appropriate upper limits for sources which do not meet significance criteria.

Next we add numerous data quality flags to aid in evaluating the reliability of the different flux measurements. We then perform a cross-match with the Spitzer/IRAC point source catalog of the GC \citep{Ramirez2008} and calculate the average astrometric accuracy of the catalog. Future work will more fully integrate the \spitzer/IRAC and other observations to enable detailed multiwavelength studies of individual sources, but here we only use the IRAC data to evaluate our astrometric accuracy. Finally, we provide an extinction estimate, including a quality assessment, for each source.  

For the final product, we are providing two versions of the catalog. The full catalog is a standard fits table consisting of 80 data columns that include all source flux measurements and uncertainties, a wide variety of measurement parameters, quality flags, and an extinction estimate. In addition, a version that includes image cutouts such as those used in this paper, is also available via an online software repository that hosts both the data and code developed as part of this effort.\footnote{\url{https://github.com/mjhankins/SFGCphotcode}} We have also created a short form of the full catalog, containing the types of measurements more typically found in point source catalogs. Details on all of the data included in both catalogs, as well as further specifics on how to access the catalogs, are provided in Appendix \ref{howto}. An example of the measurements included in the short form catalog is presented in Table \ref{short_cat}.

This catalog has been developed under the auspices of the \sofia\ Galactic Center Survey Legacy program, therefore we have sought to be consistent with the \sofia\ data products, particularly the FORCAST data cookbook.\footnote{\url{https://github.com/SOFIAObservatory/Recipes/blob/master/FORCAST_Photometry.pdf}}  As such, the catalog, and software for the catalog, have been developed using image coordinates (pixels) rather than celestial coordinates (arcseconds) throughout (1 pixel = $0.\arcsec768$). All of the photometry and accompanying measurements are derived utilizing the Astropy Project suite of Python packages \citep{astropy22} and Photutils \citep{larry_bradley_2020_4044744}.  As part of the online materials associated with this paper, we are releasing all of the code developed to create the source catalog. This is discussed in Appendix \ref{code}. Full information on how to use all codes developed for this program is also available online.\footnote{\url{https://github.com/mjhankins/SFGCphotcode}}

\subsection{Source Detection}\label{detection}
We begin the data analysis process for each field by performing background subtraction on the data. Because the GC is a very complex region with numerous extended sources like molecular clouds, we must be careful in constructing a background model that can account for large-scale variations across the image unrelated to the sources we are attempting to measure. For this purpose, we construct a 2D model background using the MMMBackground\footnote{\url{https://photutils.readthedocs.io/en/stable/api/photutils.background.MMMBackground.html}} from Photutils with a box size of 9 pixels at 25 \micron\ and 11 pixels at 37 \micron. As part of this step, we perform an initial detection of sources meeting a 3$\sigma$ significance level that also have a kernel FWHM of at least 3 pixels. Pixels meeting these criteria are masked and then the background model is constructed from the remaining emission in the image. 

Next, source detection is carried out on the background-subtracted images using two different algorithms. The first method is based on the DAOFIND algorithm \citep{Stetson87}, included in the Photutils package \citep{larry_bradley_2020_4044744} as the DAOStarFinder routine. This routine searches for local density maxima that meet a defined threshold (5$\sigma$ above the measured background of the image), and have a size and shape consistent with a 2D Gaussian kernel which we define to be similar to the FWHM of a point source at 25 and 37~\micron\ as observed by FORCAST. This method is optimal for finding point-like sources and marginally extended objects which still resemble the input Gaussian kernel; however, there are limits to how well it performs for sources that are more extended or may not appear very much like the input Gaussian kernel. 

For extended or irregular sources not well fit by the DAOStarFinder input parameters, we turn to a different method known as a Image Segmentation,\footnote{\url{https://photutils.readthedocs.io/en/stable/api/photutils.segmentation.SegmentationImage.html}} which is also implemented as part of the Photutils package. As the name suggests, image segmentation defines sources within the image as `segments' that contain a minimum number of connected pixels above the 3$\sigma$ measured background of the image, which we set to 10 pixels. An additional deblending step is used to separate sources with multiple peaks or components. Because the segmentation map has no input assumptions, in theory this enables both highly extended and irregular shaped sources to be identified and measured more accurately. The segmentation mapping routine does identify many of the compact and point sources found with DAOStarFinder. Both methods, however, have weaknesses in addition to their strengths, primarily related to their input assumptions. DAOStarFinder requires an input kernel which is best for finding sources that are similar size and shape to the kernel, in our case a 2D Gaussian kernel. This method can miss irregular sources which may not have a strong resemblance to the input kernel. Alternatively, segmentation mapping requires no input source shape, but instead relies on finding a collection of adjacent pixels above a user specified noise level. This method is good for finding extended sources or irregular shapes, however, point-like sources are typically detected at lower significance when compared to DAOStarFinder, which can result in missing some fainter sources when comparing the results of both methods. 

We therefore combine the output source lists from DAOStarFinder and segmentation mapping prior to performing any photometry.  The combined source list for each field includes a flag to indicate which method found each source. In cases where both methods find the same source, the coordinates for the source are taken from DAOStarFinder, which provides better centroiding for compact sources. 

We also examined the performance of other well known source-finding algorithms such as Photutils IRAFStarFinder,\footnote{\url{https://photutils.readthedocs.io/en/stable/api/photutils.detection.IRAFStarFinder.html}} which is based on the IRAF Starfind routine. The performance of IRAFStarFinder on the \sofia/FORCAST images was virtually identical to DAOStarFinder, consequently we do not include the IRAFStarFinder output in the final source list. 

Finally, as we examined our output lists, we found that a handful of visually identifiable sources are missed by both finding methods. This primarily occurred for sources located at the edge of fields where the dither pattern had less complete coverage or in regions with very complex background emission that limited the effectiveness of the source finding routines. In the case of \lq edge\rq\ sources, these regions are typically masked to avoid large numbers of spurious detections; however, there were several instances where this leads to missing obvious sources. Thus we interactively marked and added  these obvious but missing sources to the final lists as \lq user identified\rq. 
A final source list including the DAOStarFinder, segmentation mapping, and user identified results is created for each field and used for all subsequent photometry.

\begin{figure*}[ht]
\centering
\includegraphics[scale=0.50, ]{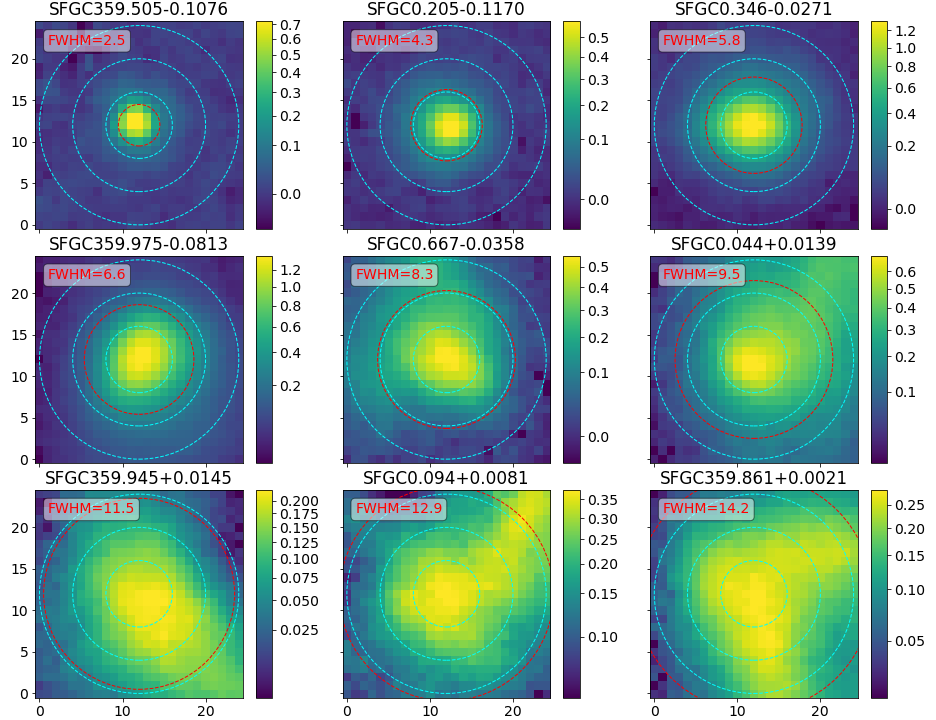}
\caption{Examples of the cutout images from the full catalog at 25 \micron.  The names correspond to those used in the SOFIA/FORCAST Galactic Center catalog. The axes are in pixels. The color scale is in Jy/pixel. The short form catalog measurements for these sources are presented in Table \ref{short_cat}.  The sources shown are good examples of point sources (upper row), compact sources (middle row), and extended sources (bottom row) included in the catalog. The teal circles are 4, 8, and 12 pixel radii apertures (3.1\arcsec, 6.1\arcsec, and 9.2\arcsec, respectively). The red circle is the FWHM value measured from the Gaussian fit, which is given in the upper left corner. \label{sources25}}
\end{figure*}

\begin{figure*}[ht]
\centering
\includegraphics[scale=0.51, ]{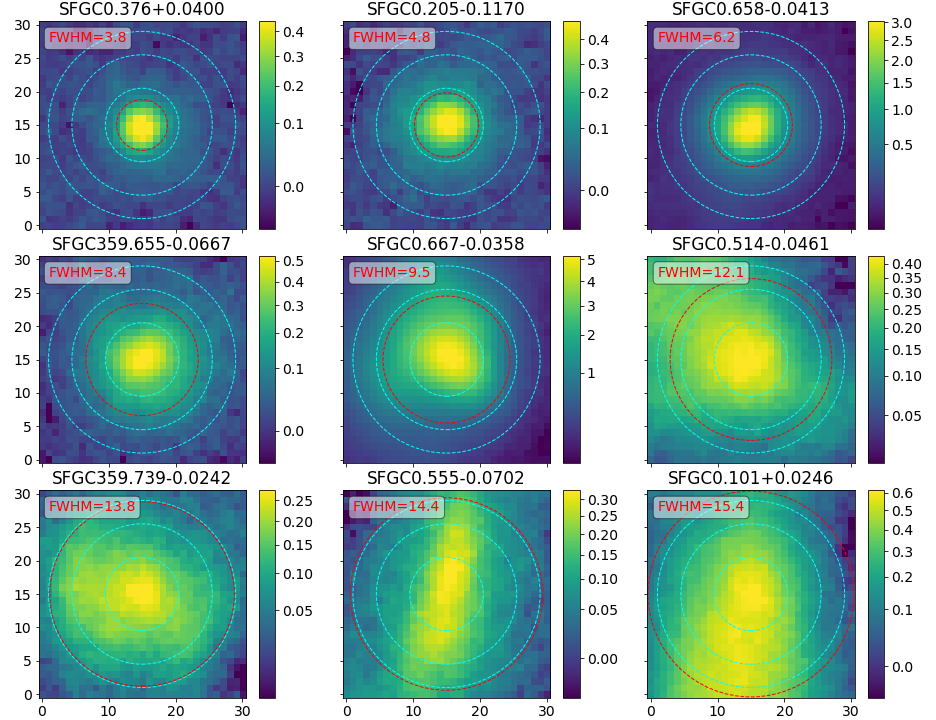}
\caption{Examples of the cutout images from the full catalog at 37 \micron.  The names correspond to those used in the SOFIA/FORCAST Galactic Center catalog. The axes are in pixels. The color scale is in Jy/pixel. The short form catalog measurements for these sources are presented in Table \ref{short_cat}.  The sources shown are good examples of point sources (upper row), compact sources (middle row), and extended sources (bottom row) included in the catalog. The teal circles are 5.5, 10.5, and 14 pixel radii apertures (4.2\arcsec, 8.0\arcsec, and 10.75\arcsec\ respectively). The red circle is the FWHM pixel value measured with the Gaussian fit, which is given in the upper left corner. The color scale is in Jy/pixel. \label{sources37}}
\end{figure*}

\subsection{Photometry}\label{photmetry}

The next stage of building the catalog is to derive fluxes for all detected sources. Figures \ref{sources25} and \ref{sources37} provide cutout images at 25~\micron\ and 37~\micron\ respectively to illustrate the diversity of sources included, and some of the challenges involved in accurately measuring the source flux given the wide variety of types we encounter.  In the figures we present good examples of the three source types we are including with confidence in the catalog data, ranging from unresolved point sources to extended sources.

The \sofia/FORCAST cookbook
discusses best practices for photometry using both aperture and 2D Moffat fitting. For aperture photometry, it is recommended to use the same apertures that are used for calibrators, a 12 pixel radius with a 15-25 pixel annulus for local background subtraction. In the Galactic Center, the combination of numerous crowded regions and regions with significant diffuse emission precludes this approach for a majority of sources. Their other recommendation is to use a slightly elongated 2D Moffat function, which allows for a better fit to the stellar wing profiles as opposed to a strict Gaussian fit, to extract the flux. This is still not sufficient for all of our sources because a significant number of the sources in the catalog are resolved compact and extended sources. 

We tested numerous iterations of aperture photometry sizes, 2D Moffat function fitting, and 2D Gaussian fitting to address the variety of sources identified in our source lists. What we found was that each method has strengths and weaknesses depending on the nature of the source. Generally, the Moffat fitting works best for sources which are unresolved. The Gaussian fitting tends to work better for compact but less circular sources. Aperture photometry is best for irregular shapes and for sources which are relatively symmetric but are resolved as compact or extended sources. 

Finally, since this catalog provides only 25 and 37 \micron\ flux values, we anticipate that a wide variety of science investigations to be addressed with the FORCAST data will require combining this catalog with observations at both longer and shorter wavelengths. The resolutions at other wavelengths, however, vary significantly. They can range from $\lesssim$1\arcsec,  typical of ground-based near-infrared observations \citep[e.g. UKIDSS,][]{UKIDSS}, to the 20$-$30\arcsec\ resolution of the {\it{Herschel}}\ 70$-$500 \micron\ observations \citep[Hi-GAL,][]{Molinari2010}; 
further demonstrating that treating all sources as if they are resolved point sources would limit the usefulness of the catalog.  

In the full form catalog we provide a best model flag which indicates which method - aperture photometry, 2D Moffat fitting, or 2D Gaussian fitting - provides the best measurement based on $\chi^2$ and other statistical analysis for each of the sources.

\subsubsection{Aperture Photometry}
Aperture photometry is the most straightforward method for deriving the source flux so we begin there. Following the FORCAST Data Handbook recommendation to perform aperture photometry on background-subtracted frames, we use the background-subtracted frames created via a process similar to that described in section \ref{detection}; however we use a larger box size (14 pixels for 25 \micron\ and 16 pixels for 37 \micron) for the background model, to avoid adversely impacting the photometry measurements for more extended sources. 

\begin{figure*}
\centering
\includegraphics[scale=0.38, ]{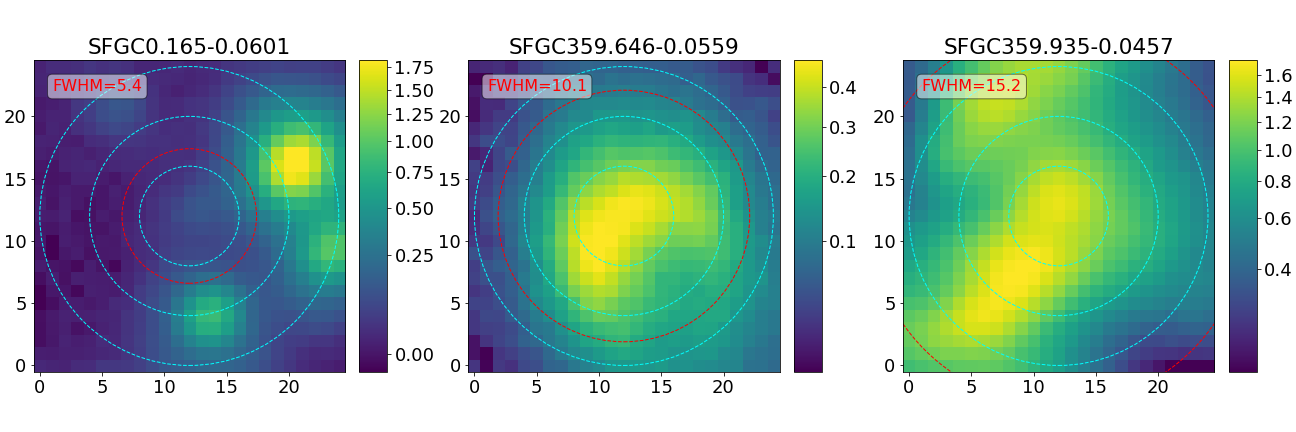}
\includegraphics[scale=0.42, ]{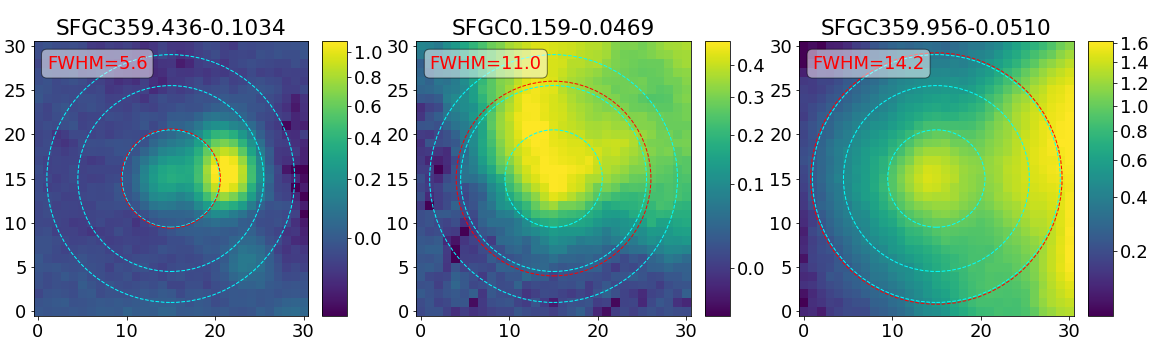}
\caption{Cutout images of irregular source objects at 25 \micron\ (Top), and 37 \micron\ (Bottom).  The overlays are the same as in Figures \ref{sources25} and \ref{sources37}, respectively.  These cutout images illustrate the challenges to deriving accurate fluxes from crowding, irregular shaped objects, and adjacent bright diffuse emission. \label{irregular}  }
\end{figure*}

We began testing source photometry using the Handbook recommendation of a 12 pixel aperture, and a 12-25 pixel annulus for local subtraction, but found that such a large aperture and background annulus did not produce the best results. Too often, within the recommended aperture or annulus there is either an additional point source or part of an extended source (see Figure \ref{irregular} for examples).  We also found we could not just use the model background subtracted image without local background-subtraction because of the complex nature of the diffuse emission. For the local background subtraction, we use the median value of the background annulus measured for each source. 

As can be seen in the variety of sources presented in Figures \ref{sources25}-\ref{sources37}, some of the sources are well isolated and appear to be typical point sources; but even for some of these we had a difficult time deriving accurate signal-to-noise values (\ref{uncertainties}).  After numerous iterations, we determined that good flux values for the full range of sources could not be obtained using only one aperture. We found that using three apertures enables us to capture meaningful fluxes for all but our most highly extended sources, and when used compared to each other, provide the user with an empirical method for exploring the nature of the source object (see Section \ref{nature}). 

In order to determine which apertures best captured both meaningful flux measurements and provided usable information regarding source type, we turned to the calibration source images taken concurrently with some of the Cycle 7 observations.  By using the diffraction profile for these bright calibration sources, we determined that the radii of the emission minima for the first, second, and third Airy Rings best met our criteria. Given the variability of the observing conditions, there were still differences in these values depending on the flight. At 25~\micron\ our best reproducible calculations of these values are 4.04$\pm$0.01, 7.8$\pm$0.2, and 11.0$\pm$1.2 pixels.  At 37~\micron\ our best calculations of these values are 5.33$\pm$0.01, 10.73$\pm$0.03, and 13.8$\pm$0.1 pixels. 

To ensure consistent and reproducible results, at 25~\micron\ we provide flux measurements using 4, 8, and 12 pixel apertures (3.1\arcsec, 6.1\arcsec, and 9.2\arcsec\ respectively). These apertures are overlaid on the 25 \micron\ source images presented in Figure \ref{sources25}. For the 37~\micron\ data we provide flux measurements for 5.5, 10.5, and 14 pixel apertures (4.2\arcsec, 8.0\arcsec, and 10.75\arcsec\ respectively). Figure \ref{sources37} provides examples of our catalog sources at 37 \micron\ overlaid with the apertures at these radii.  

Local background is subtracted using an annulus around the sources. To ensure we minimized the inclusion of source flux in our background annuli, while maintaining reproducible results, we used the same size background annulus for all of the objects and aperture measurements: 12$-$20 pixels at 25 \micron\ and 14$-$22 pixels at 37 \micron.

As can be seen in Figures \ref{sources25} and \ref{sources37}, the 8 pixel (at 25 \micron) and 10.5 pixel (at 37 \micron) apertures include nearly all of the flux for those sources which are clearly not extended, and we recommend using those flux values as the most robust for comparisons over the entire GC.  These are the aperture fluxes included in the short form catalog (Table \ref{short_cat}). 

In section \ref{analysis}, we analyze the differences in the resulting flux values and examine how the ratios of the different aperture fluxes can be used to analytically evaluate the individual sources. We also examine how the aperture results compare with the other methods we include in the full catalog as discussed below.  

\subsubsection{2D Moffat fitting}

 Since the PSF of FORCAST observations varies significantly from flight to flight, the SOFIA handbook recommends fitting sources individually, noting that the inner portions of the PSF can be well fit using a slightly elongated 2-D Moffat profile. The outer portions, however, are not well modelled by a Moffat function, with \citet{Su2017} finding that careful PSF modeling must be done to accurately measure excess emission within 10\arcsec\ of the target. Such detailed modeling, however, is beyond the scope or needs of the source catalog presented here.  We perform 2D Moffat fitting for all of the sources in our catalog as suggested in the FORCAST data cookbook,
 and do not attempt more detailed modeling. 

To determine the best-fit Moffat model for each source, we use the Levenberg–Marquardt (LevMar) least-squares fitting routine in the astropy.modeling package, along with the 2D Moffat model routine within the same package.
Although the Moffat model is expected to work well for point-like sources, our source catalogs include numerous marginally resolved and extended sources that will not be well fit by the Moffat profile. To account for these instances, we calculate a reduced $\chi^2$ statistic for each source and use this to produce a quality flag to indicate poor model fits, where the derived Moffat flux may be unreliable. In these cases where the Moffat model was unsuccessful, parameters related to the Moffat fit are masked in the final source table. 

\subsubsection{2D Gaussian fitting}
Since many of the sources are compact without necessarily being point-like, or have asymmetries which are not well handled by 2D Moffat fitting (which assumes radial symmetry), we also perform 2D Gaussian profile fitting for sources in the catalog.  In addition, in the literature, FWHM values are often provided for sources, and these values are usually derived from fits assuming a Gaussian profile. By performing and including our 2-D Gaussian fitting results in the full catalog, we are able to provide the user with comparable FWHM values for nearly all of the sources.  In Figures \ref{sources25}-\ref{irregular} we have overlaid the measured FWHM on the images.  We use this FWHM value to colormap our analysis in section \ref{analysis}.

For more extended non-symmetric sources, the largest aperture photometry currently provides the most robust measurements.

\subsubsection{Flux Uncertainties}\label{uncertainties}
Flux uncertainties quoted in the catalog are uncertainties in the photometric measurement only, and do not include calibration uncertainties.  For all \sofia/FORCAST observations the uncertainty in each pixel is dominated by a combination of sky noise and pixel response variation because the data are not ``flat-fielded" and the image background is already a residual background due to the sky subtraction that occurs as part of the observation. Following from the FORCAST Photometry cookbook,\footnote{\url{https://github.com/SOFIAObservatory/Recipes/blob/master/FORCAST-photometry\_detailed.ipynb}} we assume the uncertainty due to pixel-to-pixel response variation and the Poisson noise from the sky are approximately equal. This allows us to express the absolute measurement uncertainty, $\sigma_m$, as:
\begin{equation}
\sigma_m^2 = 2 N_{pix} \sigma_b^2  = 2 \pi r_{ap}^2 \sigma_b^2 
\end{equation}
where $N_{pix}$ is the number of pixels contained within the aperture, $r_{ap}$ is the aperture radius in pixels, and $\sigma_b$ is the standard deviation of the measured background pixels. Background measurements were obtained for each source using an annulus centered at the source location, with $r_{in}=12$ pixels and $r_{out}=20$ pixels at 25 \micron\ and $r_{in}=14$ pixels and $r_{out}=22$ pixels at 37 \micron, respectively. In the case of model photometry (Moffat2D and Gaussian 2D), we estimate the measurement uncertainty following the same method and adopt an effective aperture radius equal to $\sim1.5\times$ the measured FWHM. 

The information required to estimate the overall flux uncertainty for each source is also provided in the catalog. Following from the SOFIA/FORCAST data cookbook, the fractional flux uncertainty, $\eta$, for each source can be derived from the following expression:
\begin{equation}
    \eta^2 = (\sigma_m / F_0)^2 + (\eta_{flux})^2 +(\eta_{model})^2
\end{equation}
where $\sigma_m$ is the measured photometric uncertainty, $F_0$ is the source flux, $\eta_{flux}$ is the relative uncertainty in flux calibration, 
and $\eta_{model}$ is the relative uncertainty in the flux calibration model at the given wavelength.
The value of $\eta_{flux}$ for each source is provided in the catalog as columns named `ErrCalF25' or `ErrCalF37'. The value for $\eta_{model}$ is generally taken to be 5\% \citep{Dehaes2011}. 

\subsubsection{Shape Parameters}

The 
2-D Moffat and 2-D Gaussian fitting routines also produce parameters related to the morphological shape of the sources. For instance, the 2D Moffat parameters can be used to calculate a 1D FWHM for each source, while the 2D Gaussian parameters provide both a major and minor axis FWHM, the ratio of which is a good measure of the source asymmetry.  In the full version of the catalog, we provide the derived parameters needed to address these deviations from a point source morphology.  In the abridged version of the catalog (Table \ref{short_cat}) only the 1-D Gaussian FWHM is provided.

\subsection{Assembling the Catalog}
The catalogs are combined by first matching all of the same sources for a given wavelength, generating \lq master\rq\ tables at 25 and 37 \micron. At this stage we remove all sources with a SNR$<$3.  As we combined the tables we found that source duplication is frequent, particularly for sources that fall in the overlapping regions between two or more fields. 

Any duplicate sources are compared for quality by looking at their SNR value. The source with highest SNR is retained, with any lower SNR sources removed from the master table. The SNR provides a good discriminate for the best sources to keep since most duplicates are found near the edges of fields where the dither pattern for one observed field might not cover the source as fully as another, particularly if the source is nearer to the field center in one of the observed fields.  

Next, we match sources found at 25 and 37 \micron.  This is accomplished by doing a radial coordinate search of between the two master tables. Any sources found within 3\arcsec\ are grouped together as the same source. 

When we first combined the tables from the two different wavelengths, we identified numerous instances where a source was detected, and passed our viability tests, at one wavelength, but was {\it{not}} included in the final results at the other wavelength. The reasons were varied, but often happened because our SNR minimum was achieved at only one wavelength. In order to ensure that all the sources have measurements available from both wavelengths, we use the coordinates for the source at the wavelength already included the catalog, and perform our three aperture photometry measurements at that location at the other wavelength. We provide these values as upper limits in the full catalog.  Because of the nature of this group of sources, the 2D Moffat and 2D Gaussian measurements frequently fail, so these methods are not used to derive upper limits.  In section \ref{flags}, we discuss the quality flags included in the catalog to provide the user a measure of the reliability of the quoted fluxes.  

\subsubsection{Astrometry}\label{astrometry}

In order to estimate the astrometric uncertainty of our catalog, we cross match our source catalog with the \spitzer/IRAC catalog of the GC from \cite{Ramirez2008}. The absolute astrometry of \spitzer/IRAC is much better than SOFIA/FORCAST because of well known issues with distortion mapping across the full FoV of the FORCAST camera \citep{Adams2012}. In order to provide a reliable estimate of the uncertainties in our published source coordinates, we derive universal position uncertainties by evaluating those sources in our catalog that can be well matched to sources in the \spitzer\ catalog. 

We restrict our search of the \spitzer/IRAC catalog (henceforth SSTGC) for comparison to objects that have an 8 \micron\ counterpart to our measured sources. This is in part because the shorter wavelength bands of the SSTGC catalog typically have multiple sources within the FORCAST apertures due to the crowded nature of sources in the GC and the intrinsic resolution of FORCAST.  In addition, these shorter wavelength sources are predominantly main-sequence stars, so are not actually the same sources as those measured at the FORCAST wavelengths. The 8 \micron\ sources, on the other hand, are typically intrinsically cool enough that they are very likely the same source as the one measured in the FORCAST catalog at 25 and 37 \micron.  The SSTGC 8 \micron\ sources thereby enable us to gauge the astrometric accuracy of the coordinates in our catalog.   

\begin{figure}
\centering
\includegraphics[scale=0.41, ]{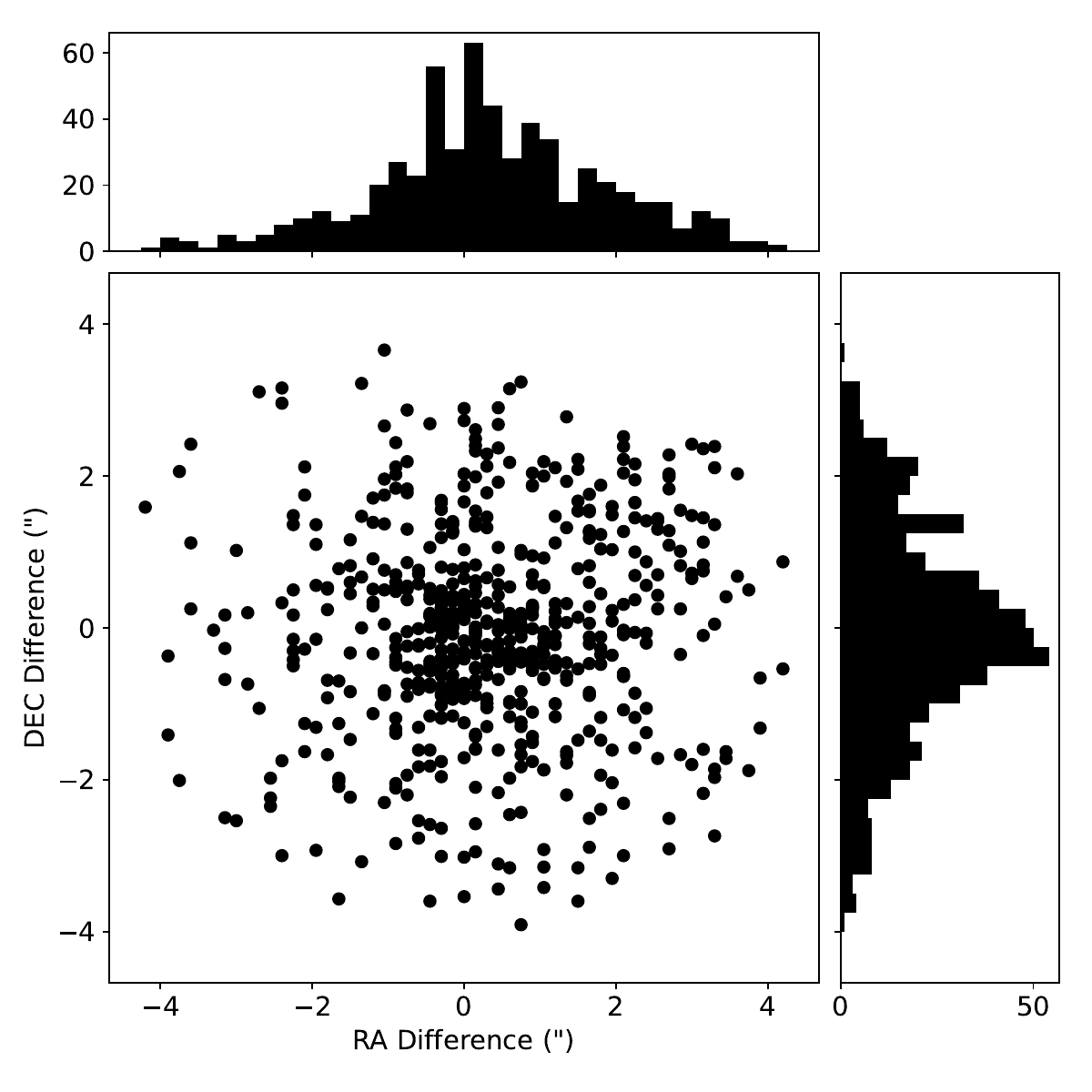}
\caption{Calculated offsets between cross-matched SOFIA/FORCAST sources in this catalog and 8 \micron\ \spitzer/IRAC sources from the SSTGC catalog \cite{Ramirez2008}. The astrometric uncertainties of the SOFIA/FORCAST GC catalog based on this analysis are RA: $+0.34\arcsec\pm1.52\arcsec$ and DEC: $-0.06\arcsec\pm1.40\arcsec$. \label{Astrometry}}
\end{figure}

After performing the source cross-match, we calculate the difference in RA and DEC from the SSTGC catalog and our catalog on a source-by-source basis. The results can be seen in figure \ref{Astrometry}. The astrometric deviation for the SOFIA/FORCAST GC catalog based on this analysis is RA: $+0.34$\arcsec$\pm$1.52$\arcsec$ and DEC: $-0.06\arcsec\pm1.40\arcsec$.  These values represent the mean of the offsets using all of the sources, with the 1$\sigma$ uncertainty derived using that mean. We note that this result is quite good considering that the FWHM calculated for ideal instrument performance of SOFIA/FORCAST is 2.8\arcsec\ at 25 \micron\ and 3.2\arcsec\ at 37 \micron. 

\begin{deluxetable*} {l|cc|rrcrc|rrcrc} 
\tabletypesize{\footnotesize}
\tablecaption{Short Form Catalog Data for sources in Figures \ref{sources25}, \ref{sources37}, and \ref{irregular}
\label{short_cat}}
\tablehead{
\colhead{}&\colhead{}&\colhead{}&\multicolumn{4}{c}{25 \micron\ Flux (Jy)} &\colhead{FWHM}&\multicolumn{4}{c}{37 \micron\ Flux (Jy)} &\colhead{FWHM} \\
\colhead{SourceID}&\colhead{RA(J2000)}&\colhead{DEC(J2000)}&\colhead{Apert.}&\colhead{$\sigma$}&\colhead{Gaussian}&\colhead{$\sigma$}&\colhead{pixels}&\colhead{Apert.}&\colhead{$\sigma$}&\colhead{Gaussian}&\colhead{$\sigma$}&\colhead{pixels}
}
\startdata
SFGC359.505-0.1076	&	17h44m51.44s	&	-29d24m52.86s	&	16.4	&	1.29	&	13.7	&	1.07	&	2.5	&	7.91	&	0.781	&	6.67	&	0.557	&	3.68	\\
SFGC0.205-0.1170	&	17h46m33.84s	&	-28d49m18.42s	&	23.1	&	1.81	&	21	&	1.65	&	4.3	&	38.1	&	3.09	&	33.5	&	2.71	&	4.76	\\
SFGC0.346-0.0271	&	17h46m32.91s	&	-28d39m15.77s	&	82.1	&	6.41	&	82.5	&	6.46	&	5.8	&	145	&	11.6	&	131	&	10.6	&	7.51	\\
SFGC359.975-0.0813	&	17h45m52.64s	&	-29d00m00.06s	&	104	&	8.11	&	109	&	8.53	&	6.6	&	144	&	11.7	&	129	&	10.5	&	8.3	\\
SFGC0.667-0.0358	&	17h47m20.35s	&	-28d23m05.07s	&	53.5	&	4.19	&	60.1	&	4.73	&	8.3	&	744	&	59.8	&	666	&	53.7	&	9.45	\\
SFGC0.044+0.0139	&	17h45m40.20s	&	-28d53m29.39s	&	63.8	&	6.23	&	76.4	&	7.41	&	9.5	&	134	&	11.7	&	114	&	10.4	&	12.4	\\
SFGC359.945+0.0145	&	17h45m25.92s	&	-28d58m32.14s	&	29	&	2.29	&	34.3	&	2.71	&	11.5	&	69	&	5.71	&	60.7	&	5.11	&	12.7	\\
SFGC0.094+0.0081	&	17h45m48.69s	&	-28d51m06.91s	&	42.5	&	4.17	&	52.8	&	5.3	&	12.9	&	86.3	&	8.79	&	75.3	&	8.12	&	13	\\
SFGC359.861+0.0021	&	17h45m16.78s	&	-29d03m14.40s	&	37.1	&	3.1	&	45.9	&	4.02	&	14.2	&	55.4	&	4.68	&	47.5	&	4.28	&	15.2	\\
SFGC0.165-0.0601	&	17h46m14.84s	&	-28d49m35.06s	&	35.6	&	2.91	&	35.9	&	3.01	&	5.4	&	56	&	4.64	&	$\cdots$	&	$\cdots$	&	$\cdots$	\\
SFGC359.646-0.0559	&	17h44m59.52s	&	-29d16m03.63s	&	55.6	&	4.36	&	65.2	&	5.15	&	10.1	&	103	&	8.39	&	89.1	&	7.3	&	11.9	\\
SFGC359.935-0.0457	&	17h45m38.61s	&	-29d00m55.53s	&	172	&	14.9	&	216	&	20.5	&	15.2	&	690	&	58.6	&	594	&	53.3	&	14.3	\\
SFGC0.376+0.0400	&	17h46m21.43s	&	-28d35m38.98s	&	2.12	&	0.236	&	1.79	&	0.148	&	2.94	&	23.1	&	1.89	&	19.5	&	1.56	&	3.8	\\
SFGC0.205-0.1170	&	17h46m33.84s	&	-28d49m18.42s	&	23.1	&	1.81	&	21	&	1.65	&	4.29	&	38.1	&	3.09	&	33.5	&	2.71	&	4.8	\\
SFGC0.658-0.0413	&	17h47m20.40s	&	-28d23m42.09s	&	31.3	&	2.45	&	30.2	&	2.36	&	5.04	&	236	&	18.9	&	213	&	17.1	&	6.2	\\
SFGC359.655-0.0667	&	17h45m03.47s	&	-29d15m53.80s	&	40.5	&	3.17	&	41.1	&	3.23	&	6.09	&	61.6	&	4.96	&	55.6	&	4.48	&	8.4	\\
SFGC0.667-0.0358	&	17h47m20.35s	&	-28d23m05.07s	&	53.5	&	4.19	&	60.1	&	4.73	&	8.3	&	744	&	59.8	&	666	&	53.7	&	9.5	\\
SFGC0.514-0.0461	&	17h47m01.10s	&	-28d31m15.52s	&	30.2	&	2.44	&	35.1	&	2.84	&	9.92	&	71.2	&	5.99	&	61.8	&	5.3	&	12.1	\\
SFGC359.739-0.0242	&	17h45m05.47s	&	-29d10m18.14s	&	21.6	&	1.7	&	25.2	&	1.99	&	9.45	&	51.9	&	4.2	&	43.8	&	3.57	&	13.8	\\
SFGC0.555-0.0702	&	17h47m12.50s	&	-28d29m55.31s	&	19.7	&	1.6	&	23	&	1.91	&	13.6	&	54.3	&	4.47	&	45.6	&	3.87	&	14.4	\\
SFGC0.101+0.0246	&	17h45m45.88s	&	-28d50m13.61s	&	48.7	&	4.79	&	60.1	&	6.13	&	14.2	&	122	&	11.6	&	103	&	10.3	&	15.4	\\
SFGC359.436-0.1034	&	17h44m40.54s	&	-29d28m15.99s	&	12.3	&	1.01	&	13.7	&	1.08	&	3.71	&	52.7	&	4.33	&	49.3	&	4.02	&	5.6	\\
SFGC0.159-0.0469	&	17h46m10.80s	&	-28d49m30.62s	&	42.8	&	3.98	&	51.7	&	4.8	&	10.4	&	82.8	&	7.68	&	71.6	&	6.84	&	11	\\
SFGC359.956-0.0510	&	17h45m42.91s	&	-28d59m59.71s	&	88.7	&	7.91	&	103	&	9.1	&	9.6	&	240	&	24.4	&	207	&	24.9	&	14.2	\\
\hline
\enddata
\tablenotetext{}{A machine readable version of the entire short form catalog can be found here: \textit{(link to contents of SFGC\_ShortCatalog.fits)}.}
\end{deluxetable*}

\subsubsection{Quality Analysis} \label{flags}
A variety of quality flags are included with the catalog to aid in evaluating individual source measurements. 

As part of the criteria used to merge the different wavelength catalogs, only those sources with SNR$\geq$3 based on the 8 pixel aperture measurements at 25 \micron\ and 10.5 pixel aperture at 37 \micron\ are retained.  However, in instances where a source is significantly detected at either 25 or 37 \micron, we perform forced photometry at the other wavelength to provide upper limit measurements.  We include a photometric quality flag at both 25 and 37 \micron\ to highlight any source aperture measurements that fail our SNR$\geq$3 significance criterion for any of the apertures.  For measurements which fail our significance criterion, we recommend using the provided $3\sigma$ upper limit measurements for the failed aperture.
 
All of the observations are background-subtracted as part of the photometric pipeline (see \ref{photmetry}), but the Galactic Center is full of complex background regions where brightness variations may not be completely subtracted off in this step, especially near molecular clouds and/or other bright complex sources (e.g. Figure \ref{irregular}). Sources with significant local background variations as measured in the source background annuli are flagged. This flag (see Table \ref{full_cat}) provides a quick way of knowing whether a source is in a relatively complex background region which could potentially impact the measured photometry.

An additional quality flag that is useful to consider is the edge flag. The edge flag is set when a source is found near the edge of a field where, because of the dither pattern, the location has been observed with less than 80\% of the exposure time obtained for the center of the field. In these regions where the dither coverage is less complete the final image noise is greater, and the observed source morphology can be effected. Thus, care should be taken when using the fluxes for sources in such regions. 

\subsubsection{Extinction} \label{extinction}
The line-of-sight extinction to the Galactic Center is notoriously large and highly variable. The flux values provided in our catalog are only measurements of the observed flux.  Since scientific investigations often require intrinsic fluxes, adjusted for the line-of-sight extinction, we are providing estimates of $\tau_{9.6}$ for each source, using values derived by \citet{Simpson2018IRS}. In that paper, she analyzed all available \spitzer/IRS observations, exploring a variety of parameters including electron densities, ionic abundances, excitation, and extinction. \citet{Simpson2018IRS} provides a detailed description of how the $\tau_{9.6}$  values we include in our catalog were derived. In brief, by using a combination of modeling of the 9.6 \micron\ silicate feature and lower limit estimates using the [\ion{S}{3}] 18.7/33.5 \micron\ lines, it is possible to estimate a $\tau_{9.6}$ value for each IRS observation.  We use these estimates based on $\tau_{9.6}$ rather than those available in the near-infrared to minimize the needed assumptions to derive absolute fluxes at 25 and 37 \micron.

The \spitzer/IRS observations used were taken from several programs and were not obtained as a comprehensive spatial survey.  Consequently, there are gaps in the coverage. To ensure we have optical depth values for all our sources, we have used a ``nearest-neighbor" interpolated $\tau_{9.6}$ map (provided by J. Simpson; private communication) to derive a $\tau_{9.6}$ value for each source based on their catalog coordinates.  We note, however, than due to the large gaps in the observations, not all $\tau_{9.6}$ values are equally reliable. Therefore, we also estimate the distance between each of our sources and the nearest IRS observed location; the larger the calculated distance from the IRS observation used, the lower the reliability of the provided $\tau_{9.6}$ value.  At distances greater than $\sim$30\arcsec\ the extinction values should be used with caution.  

We use the \citet{Chiar2006} extinction law to compute conversion factors for $\tau_{9.6}$ to $\tau_{25.2}$ and  
$\tau_{37.1}$.  We estimate these values by convolving the extinction law with the FORCAST F252 and F371 filter profiles, thereby deriving conversion factors of $\tau_{25.2}$/$\tau_{9.6}$= 0.41 and $\tau_{37.1}$/$\tau_{9.6}$=0.28.  

\subsection{Limitations for Highly Extended Sources}
In constructing this source catalog, there are sources whose derived FWHM values are $\gtrsim$12 pixels at 25 \micron\ and $\gtrsim$14 pixels at 37 \micron.  Since these sources are comparable, or even larger than, the largest extraction apertures, their catalog fluxes are probably either underestimated, or have had emission which is actually intrinsic to the source subtracted as background emission.  A future paper will supplement the full catalog provided here with additional flux measurements for these very extended sources using the segmentation mapping methods discussed above.  Inclusion of fully tested, robust values for these sources is beyond the scope of this paper. The current catalog does include an \lq very extended\rq\ flag (vExtFlag) for sources with calculated FWHM values $>$12 pixels at 25 \micron\ and $>$14 pixels at 37 \micron, whose measured fluxes, even in the largest apertures, are not reliable. We include these sources here only for completeness.
  
\begin{figure*}
\centering
\includegraphics[scale=0.44, ]{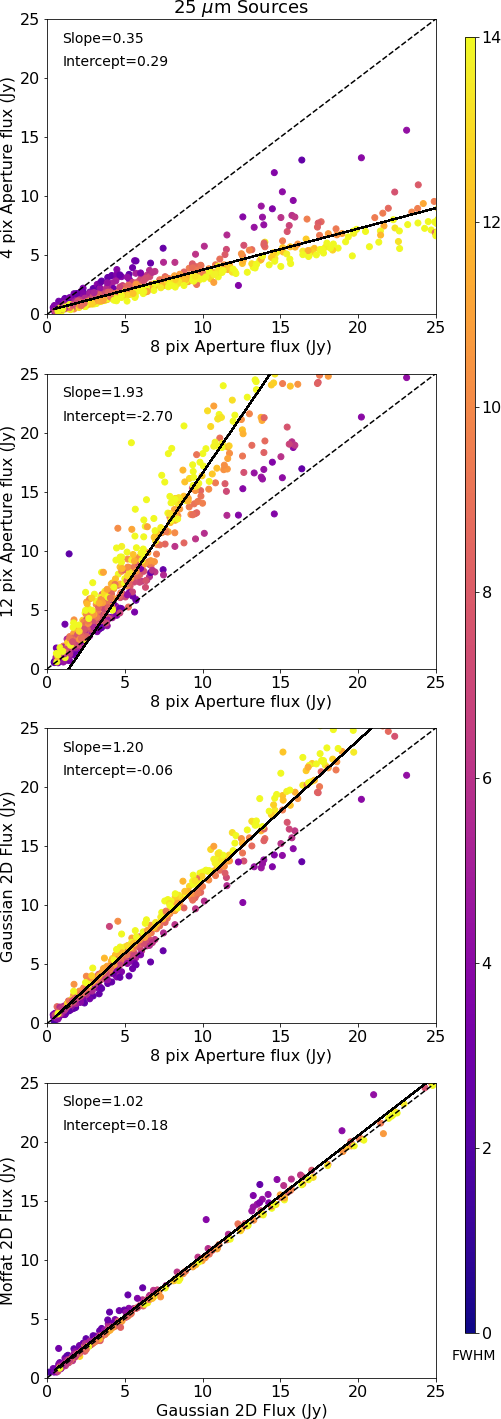}
\includegraphics[scale=0.44, ]{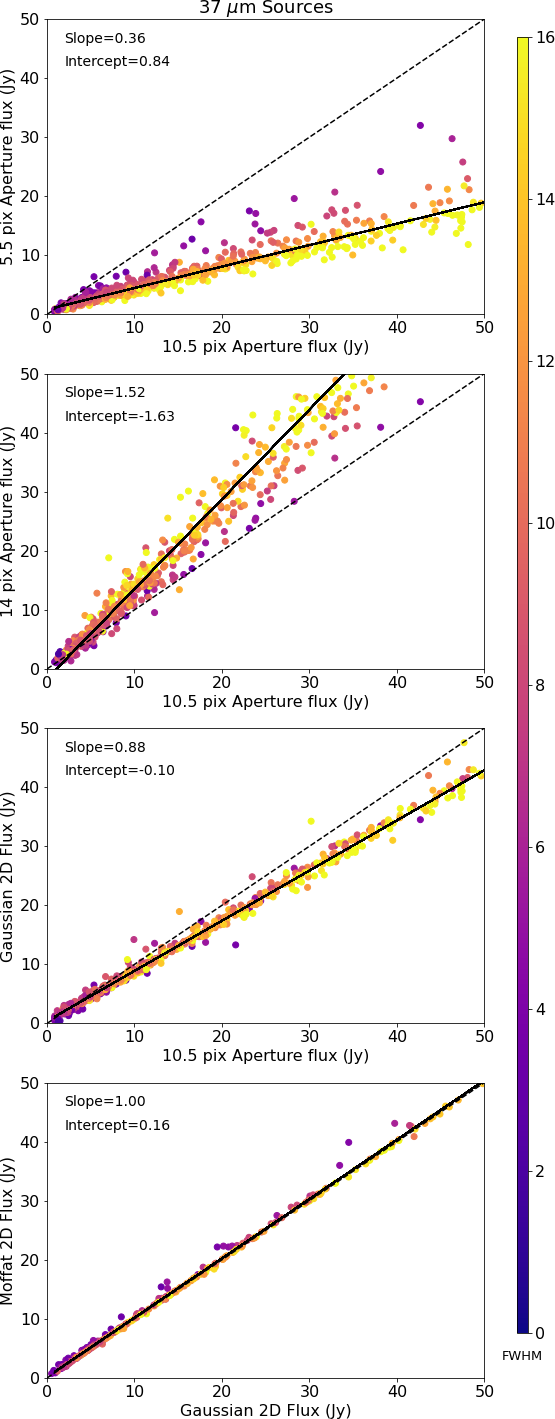}
\caption{Comparison of the fluxes as derived using the five metrics discussed in the text at 25 \micron\ (Left) and 37 \micron\ (Right). The slope and intercept values are derived using over 95\% of the sources.  The color map uses the catalog FWHM values (in pixels) derived from the 2D Gaussian fits for the sources at the respective wavelengths, ranging from 0-14 pixels (0-10.75$\arcsec$) at 25 \micron\ and 0-16 pixels 
(0-12.3$\arcsec$) at 35 \micron. \label{results}}
\end{figure*}

\section{Analysis\label{analysis}}
Given the ease with which online catalogs can now be accessed, we are producing both a long form and short form catalog.  The short form catalog provides our best source coordinates, flux measurements in Jy, uncertainties, and measured Gaussian FWHM. An example of this catalog is shown in Table \ref{short_cat}, where we list the values for the sources shown in Figures \ref{sources25}-\ref{irregular}, in the order in which they appear in the figures.  This short form catalog is more consistent with traditional point source catalogs, with only the medium aperture and best fitted fluxes included.  The long form catalog provides all of the measurements we performed and numerous quality flags, for a total of 80 columns.  This includes the five flux measurements (three apertures, 2D Moffat and 2D Gaussian fitting) per wavelength, uncertainties for each method, quality flags, shape parameters, and much more.  The full contents of this catalog are presented and discussed in detail in Appendix \ref{howto}. Figures \ref{sources25}-\ref{compactness} are made using the full catalog, which also includes the source image cutouts as shown in Figures \ref{sources25}-\ref{irregular}.

\subsection{Photometry Results}\label{photresults}
\begin{figure*}[t!]
\centering
\includegraphics[scale=0.5, ]{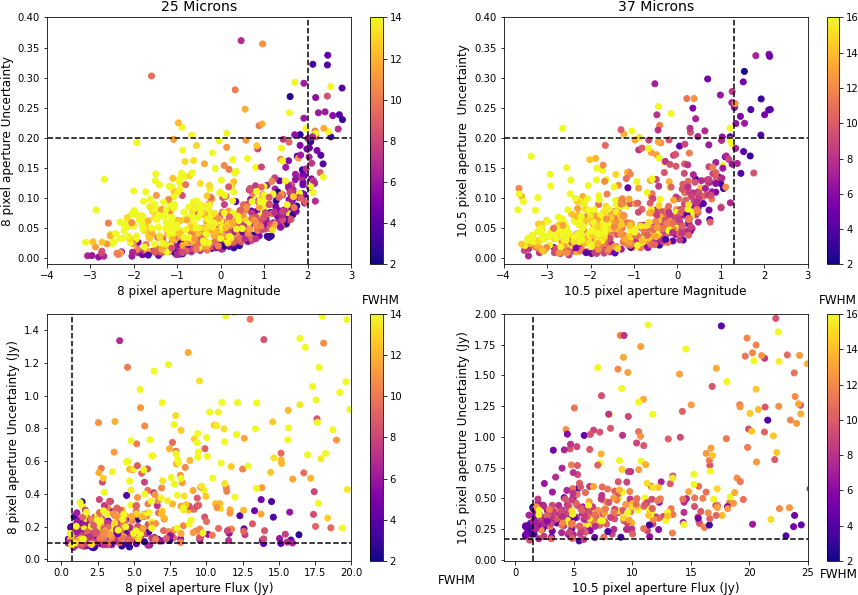}
\caption{{The measured 8 pixel and 10.5 pixel apertures fluxes and uncertainties are used to illustrate the source catalog completeness.  \it{Upper:}} The measured magnitude uncertainties versus magnitudes for the 25 \micron\ and 37 \micron\ sources.  The vertical dashed lines are at [25] = 2.0, and [37] = 1.2, the horizontal are at a magnitude uncertainty of 0.2 mag. {\it{Lower:}} The measured flux uncertainties versus measured flux.  The vertical dashed lines are for limiting fluxes of 0.75 Jy at 25 \micron, and 1.5 Jy at 37 \micron.  The baseline flux uncertainties lines are at 0.125 and 0.325 Jy respectfully. \label{completeness}}
\end{figure*}

\begin{figure*}[ht] 
\centering
\includegraphics[scale=0.42, ]{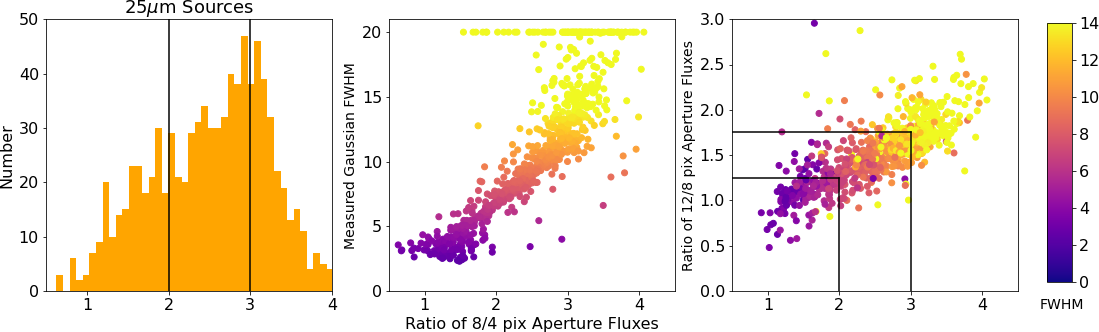}
\vspace{5pt}
\includegraphics[scale=0.42, ]{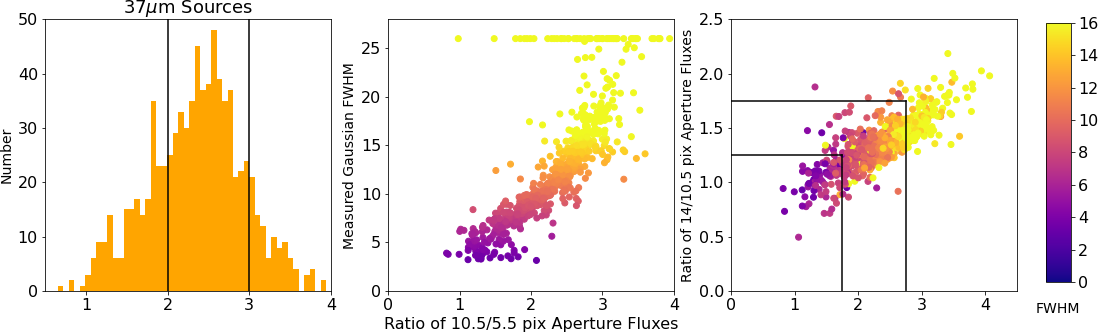}
\caption{Plots using the ratio of medium pixel aperture fluxes divided by the small pixel aperture fluxes as the x-axis to explore source types.  Top 25 \micron, and bottom 37\micron.  {\it{Left:}} The number of sources versus the aperture ratio.  {\it{Center:}} Measured FWHM from the 2D Gaussian versus the aperture ratios. The measured FWHM is unreliable beyond 20 pixels at 25 \micron, and 25 pixels at 37 \micron, therefore sources with a derived FWHM greater than these limits have been set to the limit, even though their derived aperture ratios vary. {\it{Right:}} Comparison of the large to medium apertures versus the medium to small aperture ratios.  The vertical and horizontal lines indicate the approximate ratio limits for our point source and compact source categories. \label{compactness}}
\end{figure*}

In Figure \ref{results}, we present the results from the five different photometry methods used to determine the flux in Jy at both 25 \micron\ and 37 \micron. In the figures we use the results of the medium pixel aperture measurements (8 pixels at 25 \micron\ and 10.5 pixels at 37 \micron) as our basis for comparison to fluxes derived using the smaller and larger apertures and the Gaussian 2D fitting fluxes. In addition, we compare the results from the Moffat 2D and Gaussian 2D fits. As shown in Figures \ref{sources25} and \ref{sources37}, the survey contains sources that span a large range of sizes.  In Figures \ref{results}-\ref{compactness} we use the FWHM measurements calculated by the 2D Gaussian fitting program to color map the sources, using FWHM ranges of $0-14$ pixels ($0-10.75\arcsec$) at 25 \micron\ and $0-16$ pixels ($0-12.3\arcsec$) at 35 \micron.  The color mapping enables us to visually demonstrate the effect of source size on the various derived values. 

The scatter in these plots provides a good illustration of why we are providing multiple measurement methods, rather than just one. Although we can fit sources with analytical methods, those measurements may not actually be the flux for that particular source. Figure \ref{results} also illustrates why trying to apply an simple aperture correction is not a feasible option and would actually degrade the usefulness of the results. In particular, the uncertainties involved in calculating an aperture correction would be the largest uncertainty component for any flux derived using aperture corrections.

To provide an empirical analysis of the differences between the various photometric measures, we calculate the slope and intercept values for all of comparisons of the derived fluxes shown in Figure \ref{results}.  To minimize the effect of outliers and to ensure consistency across the Figures, we use only sources with a measured medium aperture flux less than 200 Jy.  Of the 950 sources listed in the catalog, this includes a maximum of 881 sources at 25\micron\ and 829 at 37 \micron. In Figure \ref{results} we limit the displays to sources with measured fluxes less than 25 Jy at 25\micron\ and 50 Jy at 37\micron\ to better visualize the correlation between source size and measured fluxes across the various methods for the majority of the sources. Depending on the combination of photometry results illustrated in the individual plots, these limits enable us to include $\gtrsim70-80$\% of the sources. 

We find that the number of sources with reliable Gaussian and Moffat fits, as seen in Figure \ref{results}, is less than those with reliable medium aperture fluxes.  The number of sources with Gaussian fit measurements is $\sim$75\% of the medium aperture measurements at 25~\micron\ and $\sim$82\% at 37~\micron.  The number of sources with Moffat fit measurements is $\sim$50\% of the medium aperture sources at 25~\micron\ and $\sim$45\% at 37~\micron. The differences between the Gaussian and Moffat fits is likely related to the asymmetries found in the sources resulting from the elongated PSF in several of the regions due to observational difficulaties \citep{Hankins2020}. The 2D Gaussian fits are better able to handle deviations from a circularly symmetric PSF.  
As can be seen, in general, the more compact the source, as determined by their measured FWHM, the less the measured flux is affected by measurement method.  Conversely, the more extended the source, the more the derived flux depends on the chosen measurement method.  This is to be expected since the emission of an unresolved point source is by definition contained within a resolution limited radii, and should be relatively unaffected by measurement method.  The converse is true as we move to larger sources that are resolved, and therefore intrinsically extend beyond a resolution limited PSF.  

The closest matches between aperture and fitting photometry is found in the 8 pixel (25~\micron) and 10.5 pixel (37~\micron) apertures.  This is why we use those apertures as our canonical aperture photometry values.  
Interestingly, at 25~\micron\ this leads to aperture photometry values which are lower than those derived from a Gaussian fit, but gives higher values at 37 \micron. This is likely due to the differences in diffraction limits at the different wavelengths.   

The bottom row in Figure \ref{results} presents photometry from the 2D Moffat fit versus the 2D Gaussian fit.  As can be seen, they are providing essentially the same measurements regardless of source size.  This is a result of the similarities in the fitting methodologies.  

\subsection{Catalog Completeness}\label{CatCompleteness}

The observations in the FORCAST Galactic Center Legacy survey were undertaken to provide higher resolution mid-infrared observations than MSX \citep{Egan2003} in regions where the \spitzer/MIPS observations are saturated \citep{hinz2009,Hankins2020}. A consequence of this approach is that we are able to provide good measurements of the brightest sources, but at the expense of detecting and measuring faint sources.  A detailed comparison of overlapping FORCAST and MIPS sources is discussed in Section \ref{spitzer}. 

Point source catalogs at wavelengths $\lesssim$10~\micron\ typically are given in magnitudes \citep[e.g.][]{Ramirez2008}.  For those catalogs, the completeness is characterized by where the magnitude uncertainty versus magnitude curve increases rapidly. Since we anticipate that the FORCAST source catalog will be used with catalogs that use both magnitudes and fluxes, we examine the limits of the catalog using both magnitudes and fluxes.  

SOFIA/FORCAST observations are always given in flux values, so in order to investigate the catalog in magnitudes, we must first convert observed flux values to magnitudes. This has actually proven more difficult than anticipated. The FORCAST observations are solely calibrated to flux units of Jy/pixel, and there are no magnitude zero points for the FORCAST filters provided by the SOFIA Data Handbook. Communications with SOFIA support confirmed that no zero points have been measured as part of the calibration programs. Therefore we have calculated zero points for the F253 and F371 filters using the available filter functions and an analytical Vega spectrum. We calculate that at 25.3~\micron\ the zero point is 6.15 Jy, and at 37.1~\micron\ the zero point is 2.86 Jy. 

In Figure \ref{completeness}, we present both the magnitude vs. magnitude uncertainties and the flux vs. flux uncertainty plots for the canonical aperture measurements, scaled to emphasize the faint limits.  As can be seen, we find that the sensitivity limits in magnitudes are [25]  $\ga$ 2.0 and [37] $\ga$ 1.2.  In flux units the limits are 0.75 Jy at 25 \micron\ and  1.5 Jy at 37 \micron.  

In the lower right plot in Figure \ref{completeness}, the flux versus flux uncertainty at 37 \micron, we note that the measured uncertainties are consistently larger than at 25 \micron. This result is anticipated as the observations were optimized for the 25 \micron\ observations, with the 37 \micron\ taken in parallel. For the selected integration times, the time estimators indicated that the SNR ratios at 25 \micron\ should be approximately twice those at 37 \micron, which is born out in this plot. 
 
\subsection{Source Characterization\label{nature}}
The FWHM measured for all of the sources as part of the 2D Gaussian fitting photometry, although a useful metric, is not always indicative of the nature of the source.  In Figures \ref{sources25} and \ref{sources37} we presented cutout images for a very small sample of the sources to illustrate the range of roughly circular objects included in the survey, corresponding to point sources, compact sources, and extended sources.  In Figure \ref{irregular} we present additional sources at both wavelengths to illustrate repeatedly encountered anomalies.  These include, but are not limited to: faint sources near very bright emission features, irregularly shaped objects, and regions of crowding.  For these sorts of objects, the user is advised to determine whether aperture photometry or point source source fitting will provide a more accurate flux value through visual examination.  Because there are a significant number of such sources, the full online source catalog contains the cutout images used in Figures \ref{sources25}$-$\ref{irregular}, for all of the catalog sources.

\begin{figure*}[t]
\centering
\includegraphics[scale=0.45, ]{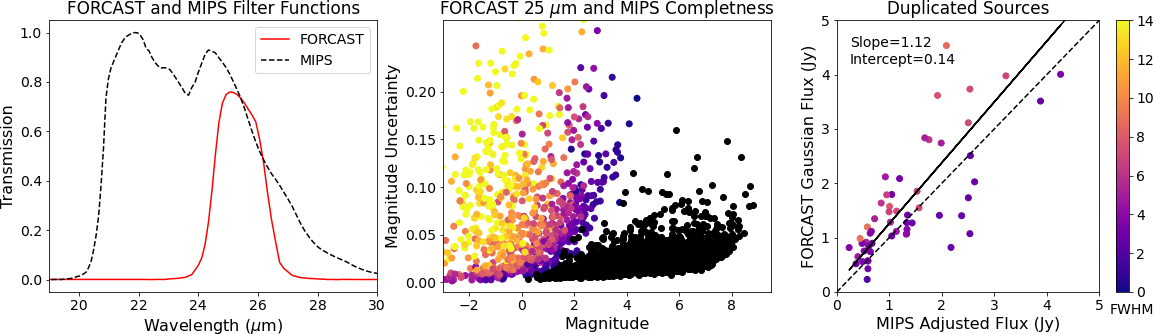} 
\caption{FORCAST at 25.3 \micron\ vs. MIPS at 23.7 \micron.  {\it{Left:}} The filter functions for the FORCAST 25 \micron\ F253 filter and the MIPS 24 \micron\ filter. We use the central wavelength of the MIPS filter to color correct the published MIPS data to the FORCAST central wavelength. The significant differences in the filter functions highlights why comparisons using the adjusted fluxes are for illustration purposes only. {\it{Middle:}} Magnitude vs. Magnitude uncertainties using the FORCAST 2D Gaussian source fits and the adjusted MIPS fluxes.  {\it{Right:}} FORCAST vs. MIPS measurements for the 58 matched sources.  In general there is good agreement, but there is significant scatter resulting from the accuracy of measurements for the duplicate sources.  \label{MIPS}}
\end{figure*}

To examine the catalog sources in aggregate, We have found that by using a combination of the measured FWHM  and ratios of the fluxes in different size apertures, we are able to roughly separate the sources into our four categories of point soures, compact sources, extended sources, and highly extended sources. In Figure \ref{compactness} we demonstrate how these metrics can be used to investigate the population in general.  In Figure \ref{compactness} we evaluate the 8/4 pixel aperture ratio at 25 \micron\ and the 10.5/5.5 pixesl aperture ratio at 37 \micron\ by number, correlation to the measured FWHM and correlation the 12/8 pixel aperture ratio at 25 \micron\ and the 14/10.5 pixel aperture ratio at 37 \micron\ to differentiate source type.  
Interestingly, as can be seen in Figure \ref{compactness}, at 25 \micron, there appear to be two peaks of the 8/4 aperture flux ratio around 2.0 and 3.0. These are the approximate values which separate point from compact and compact from extended sources.  By contrast, the ratio at 37 \micron\ shows a single peak at approximately 2.5. One interpretation of this result is that due to the higher resolution and signal-to-noise at 25 \micron, the aperture photometry is more readily able to separate unresolved sources from more resolvable extended sources, distinctions which are more difficult at 37 \micron.    

If we plot the measured FWHM versus the aperture ratios, what we see is general agreement between aperture ratios and the measured FWHM, with a gradual and continuous correlation seen as we move to from compact to extended sources.  There are however, a number of sources whose 8/4 pixel (or 10.5/5.5 pixel) flux ratios indicate they are less point like, resulting in larger aperture ratio values. The divergence in these values is due to difference aspects of crowding, either including additional point sources within the aperture radii (e.g. SFGC0.165-0.0601 in Figure \ref{irregular}), or nearby diffuse emission  (e.g. SFGC359.956-0.0510 in Figure \ref{irregular}).

Finally, we find that if we plot the large/medium versus medium/small apertures, we are best able differentiate source types. The lines shown in the right most plots in Figure \ref{compactness}, correspond to the ratios separating our different source types. It is to be noted that these ratio limits are consistent with the source types presented in Figures \ref{sources25} and \ref{sources37}.  As can be seen, there are sources which are exceptions to all of these trends clearly visible in the plot.  The combination of these observed trends, the anomalies, and the wide variety of source types are the motivation for providing the user with a wide variety of measurements and making cutout images available for all of the sources.

\subsection{Comparison of FORCAST and MIPS catalogs}\label{spitzer}
Although the FORCAST Galactic center survey targeted regions saturated in the MIPS survey, there is some overlap with unsaturated regions containing sources included in the MIPS point source catalog.  The filters used in the MIPS 24 \micron\ observations and the FORCAST F253 filter are, however, significantly different (see Figure \ref{MIPS}).  In order to accurately compare the overlapping sources, we must first adjust the derived fluxes.  We do this here by using a simple color correction calculated using the filter profiles and a Vega spectrum which was used to adjust the MIPS (23.7 \micron) fluxes to the FORCAST (25.3 \micron) values.

Using the adjusted MIPS magnitudes, we revisit the magnitude versus magnitude uncertainty plot at 25 \micron, adding in the MIPS sources located within a rectangular region defined by the boundaries of the FORCAST data (see Figure \ref{fig1}). As we can see in Figure \ref{MIPS}, the FORCAST data is first and foremost a bright source complement to the MIPS survey.  We use the FORCAST 2D Gaussian fitted magnitudes for this plot as they are the most similar of our five photometry methods to the TinyTim model PSF fitting method used in the MIPS survey catalog \citep{hinz2009,Krist2011}.  As can be seen, the overlapping magnitudes are limited to [25] $\sim 0-3$.  Unfortunately, this is also the region in which the measurement uncertainties for the FORCAST observations are rapidly increasing.  

\begin{figure*}[t!]
\centering
\includegraphics[scale=0.58, ]{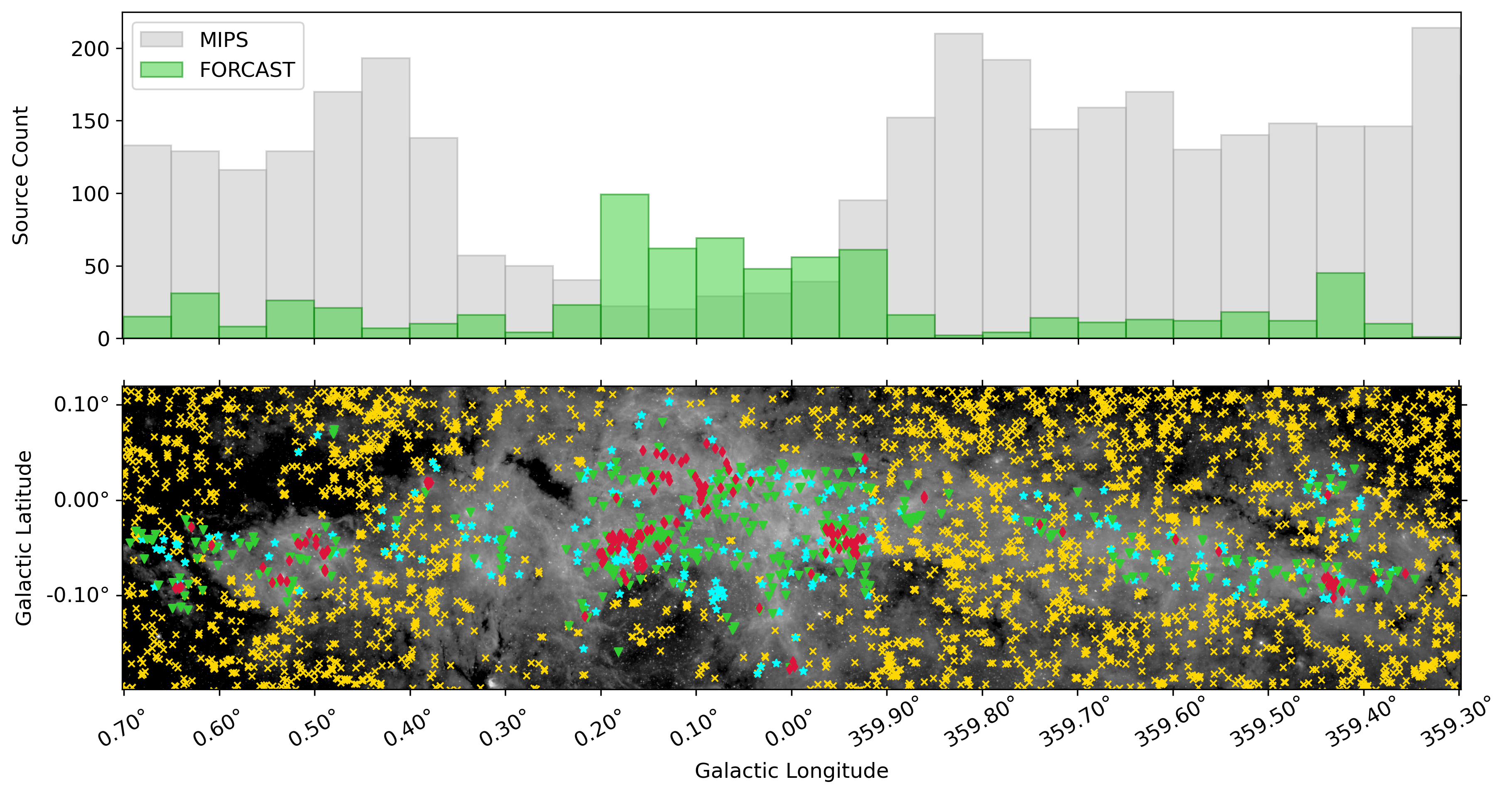} 
\caption{Spatial distribution of the 25 \micron\ sources in the \sofia/FORCAST catalog and the \spitzer/MIPS catalog \citep{hinz2009}. \textit{Top:} A histogram of source counts as a function of Galactic longitude. \textit{Lower:} FORCAST sources that are nominal point sources (FWHM $<$ 6 pix) are marked as blue stars; compact sources (6 $<$ FWHM $<$12 pix) are marked as green triangles;  extended sources (12 $<$ FWHM $<$ 18 pix) are marked as red diamonds.  The MIPS sources are represented as yellow crosses, and the background image is a greyscale \spitzer/IRAC 8 \micron\ map of the region displayed in a log-stretch. As can be seen, the \sofia/FORCAST catalog very effectively fills in the regions missed by MIPS.  Clustered extended sources highlight Sgr B1, QPS, the Arches, Sgr A, and Sgr C as labeled in Figure \ref{fig1}.  \label{distribution} }
\end{figure*}

If we set a 2\arcsec\ matching radius and search for sources identified in both surveys, we find only 58 duplicates. These are plotted in the final panel of Figure \ref{MIPS}.  As can be seen, there is a good general agreement for these sources, although there is significant scatter.  In the middle plot we can see the reasons for both the paucity of duplicate sources and the scatter.  There is very little overlap in magnitude space (and physical space, see Figure \ref{distribution}), even if we look at all of the sources in the central $\sim$200 parsecs; this was actually by design. Another prominent source of uncertainty when comparing the MIPS and FORCAST fluxes comes in the color correction due to differences in the filter profiles (Figure \ref{MIPS}, left). Notably, the MIPS 24 \micron\ filter is much bluer than the FORCAST 25 \micron\ filter, and accurate correction for these differences requires applying knowledge of the individual source spectra which is beyond the scope of this work. Considering these factors, we are unable to provide a meaningful measurement of universal differences between the observed fluxes in the two catalogs. However, this analysis clearly shows the difference in phase space occupied by the majority of FORCAST sources compared to MIPS, most of which was part of the survey design. 

In Figure \ref{distribution} we present the spatial distribution for the \sofia/FORCAST catalog sources and the MIPS catalog sources. The FORCAST sources are color coded using our point, compact, and extended source types following the designations discussed in section \ref{nature}.  We overlay the figure with the sources from the 24 \micron\ \spitzer/MIPS catalog surrounding the full FORCAST observations. Comparing the distribution of sources between the catalogs, we first note that the \sofia/FORCAST sources successfully fill in the regions that do not have sources included in the MIPS source catalog due to saturation. We further note that the sources identified as extended in our catalog cluster where one would expect them, highlighting Sgr B1, the Sickle, the Arches, Sgr A, and Sgr C as labeled in Figure \ref{fig1}. Regions with relatively few sources appear to be coincident with known molecular clouds such as the `Brick' near $l\sim0.25\degree$ and the `dust ridge' feature \citep{Lis1994,Immer2012DustRidge} which extends to near Sgr B.   

\subsection{Preview of Stellar Population Analysis \label{cmd}}

\begin{figure*}[t]
\centering
\includegraphics[scale=0.53, ]{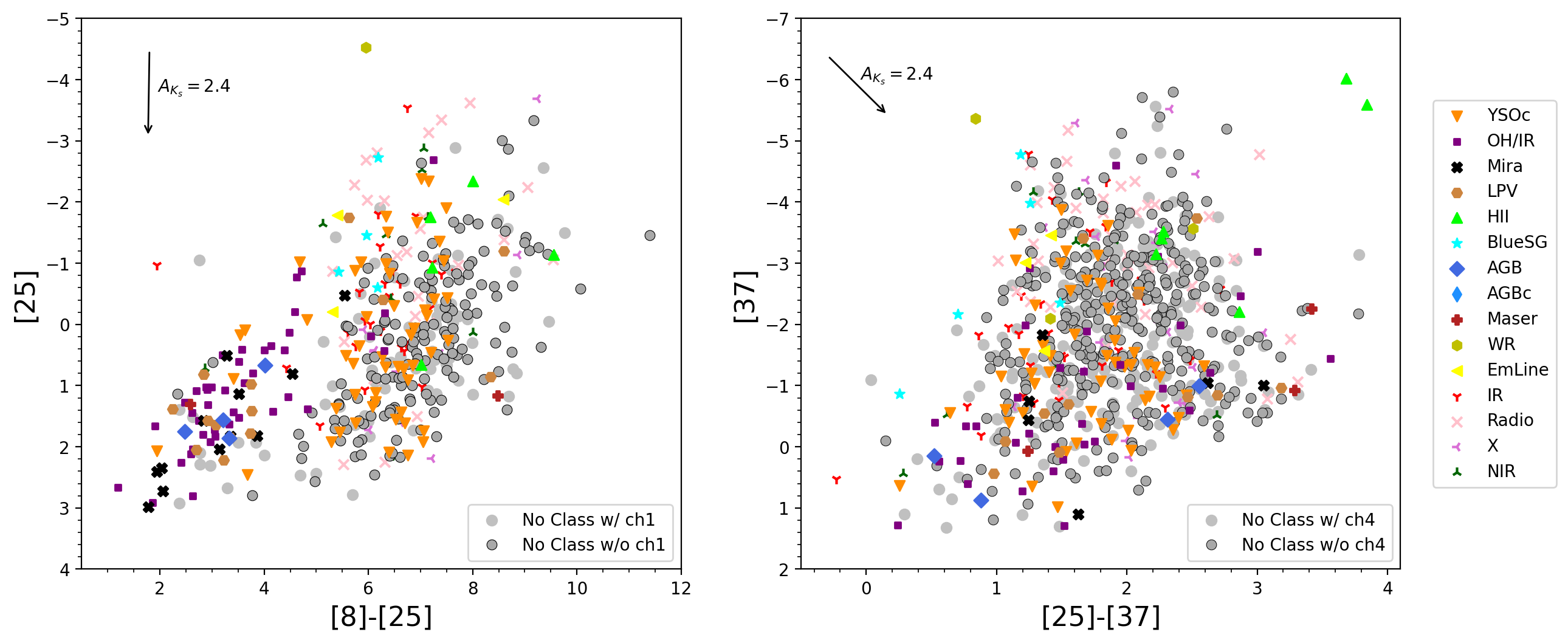} 
\caption{\textit{Left:} [8]-[25] vs. [25] color-magnitude diagram produced using data from the SSTGC catalog \citep{Ramirez2008} as well as data from the catalog presented in this paper. \textit{Right:} [37] vs. [25]-[37] color-magnitude diagram. Both panels show a reddening vector consistent with extinction measurements for the GC \citep[$A_{K_s}=2.4$;][]{Fritz2011}. For these plots, sources in the catalog have been cross matched against known objects in SIMBAD \citep{SIMBAD2000}, and any source type that has 3 or more instances has been labeled. We note there is a wide variety of source types identified from the literature; however, more than half of the objects in the catalog do not have a known object type at present. These unclassified sources are labeled `No Class' and are represented as grey circles in the plots. Additional information on the `No Class' sources is provided to indicate whether they have either \spitzer/IRAC channel 1 (3.6 \micron) or channel 4 (8.0 \micron) counterparts in the SSTGC catalog.\label{CMD} }
\end{figure*}

While a complete and detailed study of the sources contained in this catalog is reserved for future papers, we provide here a first look at the parameter space that can be explored with the catalog. We present two color-magnitude diagrams (CMDs) in Figure \ref{CMD}. As part of this analysis, a cross-match of the catalog was performed with the SIMBAD database \citep{SIMBAD2000} in order to identify sources that have been observed and classified in the literature.  We use the 8~\micron\ magnitudes from the SSTGC, matched in our astrometric analysis (section \ref{astrometry}), and our derived 25~\micron\ magnitudes (section \ref{CatCompleteness}) to create a [25] vs. [8]$-$[25] CMD (Figure \ref{CMD}).  We find distinctive clustering by source types with OH/IR, AGB, and Mira variables being significantly less red than known YSO candidates (YSOc). This clustering of sources using the FORCAST catalog is similar to what is observed in comparing \spitzer/IRAC 8~\micron~ data and \spitzer/MIPS 24~\micron~ data in other regions of sky that have quality MIPS 24~\micron~ data \citep[e.g. the LMC;][]{Whitney2008}. In addition, in the region of previously identified YSO candidates, we find numerous additional sources included in the \sofia/FORCAST catalog that have yet to be classified.  Examining these sources with the latest models for Massive YSOs (MYSOs) \citep[e.g.][]{Fedriani2023}, is the primary objective for the next phase of the \sofia/FORCAST Galactic Center Legacy program.   

We also present a CMD using just the \sofia/FORCAST 25~\micron\ and 37~\micron\ data to explore the potential for newly discovered sources.  Although these filters are not as separated in wavelength space as the 8~\micron\ and 25~\micron\ observations, we are able to show that there are many more sources included in the catalog than just those with a corresponding 8~\micron\ counterpart.  These sources are potentially interesting as they would appear to be much redder than what is found in the SSTGC catalog and represent previously unexplored parameter space. 

Finally, we note that a majority (65\%) of the catalog sources do not have a known classification in SIMBAD, and many do not have a well defined object type, but are rather labeled by the wavelength space in which they have been identified such as radio, x-ray, infrared, or near-infrared. These two CMDs clearly illustrate that there is great potential for identifying other objects of significant interest among these less well studied sources. This catalog has been developed to provide a versatile aid for future investigations of these sources.

\section{Summary}
We present here the source catalog for all of the fields observed with SOFIA/FORCAST in the Galactic Center at 25 \micron\ (F253) and 37 \micron\ (F371). The catalog is available for download on {\it{IRSA}}, and comes in two forms: an abbreviated version which is analogous to a typical point source catalog (see Table \ref{short_cat}), and the full version described in detail in Appendix \ref{howto}. For a typical use case comparing source fluxes over the entire catalog, we recommend use of the 8 pixel aperture measurements at 25 \micron\ and the 10.5 pixel aperture measurements at 37 \micron\ similar to what is presented in section \ref{cmd}. However, it may be wise to take into account the measured FWHM given the varied nature of sources in the catalog. These measurements are provided in both the short form catalog discussed above, and the full catalog presented in Appendix \ref{howto}.

In certain cases, a user may still need additional measurements which are provided in the full catalog. The full catalog contains five measurements of the flux for each identified source: 3 different aperture radii measurements, and fitting using 2D Moffat and 2D Gaussian methods. Additionally, there are quality flags and shape measurements to aid in addressing the quality of sources and their derived fluxes. Flux upper limits are provided for sources that do not meet significance criteria for the catalog. In addition to flux and fit measurements, the FITS table version of the catalog hosted with the code on Zenodo contains cutout images, such as those presented in Figures \ref{sources25}$-$\ref{irregular}, for all of the sources.

Further work on the \sofia/FORCAST Galactic Center Legacy program is currently underway. As noted earlier in the text, we are working to develop a supplementary catalog for highly extended sources, which the measurements presented here are likely only lower limits of the true source fluxes. The next phase of science analysis for this program is focused on studying stellar populations contained within the region, and in particular YSOs. An overarching goal of the legacy program is to provide a measure for the star formation rate of GC. The catalog presented in this paper is an important step towards that goal, though we note that this catalog can contribute to various types of studies within this interesting and important region within our galaxy.

\vspace{3mm}
\emph{Acknowledgments:} 
This work is based on observations made with the NASA/DLR Stratospheric Observatory for Infrared Astronomy (SOFIA). SOFIA science mission operations are conducted jointly by the Universities Space Research Association, Inc. (USRA), under NASA contract NNA17BF53C, and the Deutsches SOFIA Institut (DSI) under DLR contract 50 OK 0901. Financial support for this work was provided by NASA through SOFIA Awards 07$\_$0189 and 09$\_$0216 issued by USRA under NASA contract NNA17BF53C. This work was also supported by Arkansas Space Grant Consortium under NASA Training Grant 80NSSC20M0106. Additionally, C.\ Battersby  gratefully  acknowledges  funding  from  National  Science  Foundation  under  Award  Nos. 1816715, 2108938, 2206510, and CAREER 2145689, as well as from the National Aeronautics and Space Administration through the Astrophysics Data Analysis Program under Award No. 21-ADAP21-0179 and through the SOFIA archival research program under Award No. 09$\_$0540. 

We would also like to thank the many individuals from the USRA Science and Mission Ops teams that made the SOFIA/FORCAST Galactic Center Survey possible. 

This work made use of data products from the Spitzer Space Telescope, which is operated by the Jet Propulsion Laboratory, California Institute of Technology, under a contract with NASA. This research has also made use of the NASA/IPAC Infrared Science Archive, which is operated by the Jet Propulsion Laboratory, California Institute of Technology, under contract with the National Aeronautics and Space Administration. Finally, this research has also made use of the VizieR catalog access tool, CDS, Strasbourg, France. The original description of the VizieR service was published in A\&AS 143, 23.This research has also made use of the SIMBAD database, operated at CDS, Strasbourg, France 

\facilities{SOFIA, Spitzer}

\software{Astropy \citep{astropy2013,astropy2018,astropy22}, 
Photutils \citep{larry_bradley_2020_4044744}, SAOds9 \citep{SAOds92000,JoyeMandel2003}, REDUX \citep{Clarke2015},SIMBAD \citep{SIMBAD2000}, VizieR \citep{vizier} }

\bibliography{main}{}

\begin{thebibliography}{}
\expandafter\ifx\csname natexlab\endcsname\relax\def\natexlab#1{#1}\fi
\providecommand{\url}[1]{\href{#1}{#1}}
\providecommand{\dodoi}[1]{doi:~\href{http://doi.org/#1}{\nolinkurl{#1}}}
\providecommand{\doeprint}[1]{\href{http://ascl.net/#1}{\nolinkurl{http://ascl.net/#1}}}
\providecommand{\doarXiv}[1]{\href{https://arxiv.org/abs/#1}{\nolinkurl{https://arxiv.org/abs/#1}}}

\bibitem[{{Adams} {et~al.}(2012){Adams}, {Herter}, {Gull}, {Schoenwald}, {Henderson}, {Keller}, {De Buizer}, {Stacey}, {Nikola}, {Vacca}, {Hirsch}, {Wang}, \& {Helton}}]{Adams2012}
{Adams}, J.~D., {Herter}, T.~L., {Gull}, G.~E., {et~al.} 2012, in Society of Photo-Optical Instrumentation Engineers (SPIE) Conference Series, Vol. 8446, Ground-based and Airborne Instrumentation for Astronomy IV, ed. I.~S. {McLean}, S.~K. {Ramsay}, \& H.~{Takami}, 844616, \dodoi{10.1117/12.926600}

\bibitem[{{Astropy Collaboration} {et~al.}(2013){Astropy Collaboration}, {Robitaille}, {Tollerud}, {Greenfield}, {Droettboom}, {Bray}, {Aldcroft}, {Davis}, {Ginsburg}, {Price-Whelan}, {Kerzendorf}, {Conley}, {Crighton}, {Barbary}, {Muna}, {Ferguson}, {Grollier}, {Parikh}, {Nair}, {Unther}, {Deil}, {Woillez}, {Conseil}, {Kramer}, {Turner}, {Singer}, {Fox}, {Weaver}, {Zabalza}, {Edwards}, {Azalee Bostroem}, {Burke}, {Casey}, {Crawford}, {Dencheva}, {Ely}, {Jenness}, {Labrie}, {Lim}, {Pierfederici}, {Pontzen}, {Ptak}, {Refsdal}, {Servillat}, \& {Streicher}}]{astropy2013}
{Astropy Collaboration}, {Robitaille}, T.~P., {Tollerud}, E.~J., {et~al.} 2013, \aap, 558, A33, \dodoi{10.1051/0004-6361/201322068}

\bibitem[{{Astropy Collaboration} {et~al.}(2018){Astropy Collaboration}, {Price-Whelan}, {Sip{\H{o}}cz}, {G{\"u}nther}, {Lim}, {Crawford}, {Conseil}, {Shupe}, {Craig}, {Dencheva}, {Ginsburg}, {VanderPlas}, {Bradley}, {P{\'e}rez-Su{\'a}rez}, {de Val-Borro}, {Aldcroft}, {Cruz}, {Robitaille}, {Tollerud}, {Ardelean}, {Babej}, {Bach}, {Bachetti}, {Bakanov}, {Bamford}, {Barentsen}, {Barmby}, {Baumbach}, {Berry}, {Biscani}, {Boquien}, {Bostroem}, {Bouma}, {Brammer}, {Bray}, {Breytenbach}, {Buddelmeijer}, {Burke}, {Calderone}, {Cano Rodr{\'\i}guez}, {Cara}, {Cardoso}, {Cheedella}, {Copin}, {Corrales}, {Crichton}, {D'Avella}, {Deil}, {Depagne}, {Dietrich}, {Donath}, {Droettboom}, {Earl}, {Erben}, {Fabbro}, {Ferreira}, {Finethy}, {Fox}, {Garrison}, {Gibbons}, {Goldstein}, {Gommers}, {Greco}, {Greenfield}, {Groener}, {Grollier}, {Hagen}, {Hirst}, {Homeier}, {Horton}, {Hosseinzadeh}, {Hu}, {Hunkeler}, {Ivezi{\'c}}, {Jain}, {Jenness}, {Kanarek}, {Kendrew}, {Kern}, {Kerzendorf}, {Khvalko}, {King}, {Kirkby}, {Kulkarni},
  {Kumar}, {Lee}, {Lenz}, {Littlefair}, {Ma}, {Macleod}, {Mastropietro}, {McCully}, {Montagnac}, {Morris}, {Mueller}, {Mumford}, {Muna}, {Murphy}, {Nelson}, {Nguyen}, {Ninan}, {N{\"o}the}, {Ogaz}, {Oh}, {Parejko}, {Parley}, {Pascual}, {Patil}, {Patil}, {Plunkett}, {Prochaska}, {Rastogi}, {Reddy Janga}, {Sabater}, {Sakurikar}, {Seifert}, {Sherbert}, {Sherwood-Taylor}, {Shih}, {Sick}, {Silbiger}, {Singanamalla}, {Singer}, {Sladen}, {Sooley}, {Sornarajah}, {Streicher}, {Teuben}, {Thomas}, {Tremblay}, {Turner}, {Terr{\'o}n}, {van Kerkwijk}, {de la Vega}, {Watkins}, {Weaver}, {Whitmore}, {Woillez}, {Zabalza}, \& {Astropy Contributors}}]{astropy2018}
{Astropy Collaboration}, {Price-Whelan}, A.~M., {Sip{\H{o}}cz}, B.~M., {et~al.} 2018, \aj, 156, 123, \dodoi{10.3847/1538-3881/aabc4f}

\bibitem[{{Astropy Collaboration} {et~al.}(2022){Astropy Collaboration}, {Price-Whelan}, {Lim}, {Earl}, {Starkman}, {Bradley}, {Shupe}, {Patil}, {Corrales}, {Brasseur}, {N{\"o}the}, {Donath}, {Tollerud}, {Morris}, {Ginsburg}, {Vaher}, {Weaver}, {Tocknell}, {Jamieson}, {van Kerkwijk}, {Robitaille}, {Merry}, {Bachetti}, {G{\"u}nther}, {Aldcroft}, {Alvarado-Montes}, {Archibald}, {B{\'o}di}, {Bapat}, {Barentsen}, {Baz{\'a}n}, {Biswas}, {Boquien}, {Burke}, {Cara}, {Cara}, {Conroy}, {Conseil}, {Craig}, {Cross}, {Cruz}, {D'Eugenio}, {Dencheva}, {Devillepoix}, {Dietrich}, {Eigenbrot}, {Erben}, {Ferreira}, {Foreman-Mackey}, {Fox}, {Freij}, {Garg}, {Geda}, {Glattly}, {Gondhalekar}, {Gordon}, {Grant}, {Greenfield}, {Groener}, {Guest}, {Gurovich}, {Handberg}, {Hart}, {Hatfield-Dodds}, {Homeier}, {Hosseinzadeh}, {Jenness}, {Jones}, {Joseph}, {Kalmbach}, {Karamehmetoglu}, {Ka{\l}uszy{\'n}ski}, {Kelley}, {Kern}, {Kerzendorf}, {Koch}, {Kulumani}, {Lee}, {Ly}, {Ma}, {MacBride}, {Maljaars}, {Muna}, {Murphy}, {Norman},
  {O'Steen}, {Oman}, {Pacifici}, {Pascual}, {Pascual-Granado}, {Patil}, {Perren}, {Pickering}, {Rastogi}, {Roulston}, {Ryan}, {Rykoff}, {Sabater}, {Sakurikar}, {Salgado}, {Sanghi}, {Saunders}, {Savchenko}, {Schwardt}, {Seifert-Eckert}, {Shih}, {Jain}, {Shukla}, {Sick}, {Simpson}, {Singanamalla}, {Singer}, {Singhal}, {Sinha}, {Sip{\H{o}}cz}, {Spitler}, {Stansby}, {Streicher}, {{\v{S}}umak}, {Swinbank}, {Taranu}, {Tewary}, {Tremblay}, {de Val-Borro}, {Van Kooten}, {Vasovi{\'c}}, {Verma}, {de Miranda Cardoso}, {Williams}, {Wilson}, {Winkel}, {Wood-Vasey}, {Xue}, {Yoachim}, {Zhang}, {Zonca}, \& {Astropy Project Contributors}}]{astropy22}
{Astropy Collaboration}, {Price-Whelan}, A.~M., {Lim}, P.~L., {et~al.} 2022, \apj, 935, 167, \dodoi{10.3847/1538-4357/ac7c74}

\bibitem[{{Barnes} {et~al.}(2017){Barnes}, {Longmore}, {Battersby}, {Bally}, {Kruijssen}, {Henshaw}, \& {Walker}}]{Barnes2017}
{Barnes}, A.~T., {Longmore}, S.~N., {Battersby}, C., {et~al.} 2017, \mnras, 469, 2263, \dodoi{10.1093/mnras/stx941}

\bibitem[{Bradley {et~al.}(2020)Bradley, Sipőcz, Robitaille, Tollerud, Vinícius, Deil, Barbary, Wilson, Busko, Günther, Cara, Conseil, Bostroem, Droettboom, Bray, Bratholm, Lim, Barentsen, Craig, Pascual, Perren, Greco, Donath, de~Val-Borro, Kerzendorf, Bach, Weaver, D'Eugenio, Souchereau, \& Ferreira}]{larry_bradley_2020_4044744}
Bradley, L., Sipőcz, B., Robitaille, T., {et~al.} 2020, astropy/photutils: 1.0.0, 1.0.0,  Zenodo, \dodoi{10.5281/zenodo.4044744}

\bibitem[{{Bryant} \& {Krabbe}(2021)}]{Bryant2021}
{Bryant}, A., \& {Krabbe}, A. 2021, \nar, 93, 101630, \dodoi{10.1016/j.newar.2021.101630}

\bibitem[{{Chiar} \& {Tielens}(2006)}]{Chiar2006}
{Chiar}, J.~E., \& {Tielens}, A.~G.~G.~M. 2006, \apj, 637, 774, \dodoi{10.1086/498406}

\bibitem[{{Clarke} {et~al.}(2015){Clarke}, {Vacca}, \& {Shuping}}]{Clarke2015}
{Clarke}, M., {Vacca}, W.~D., \& {Shuping}, R.~Y. 2015, Astronomical Society of the Pacific Conference Series, Vol. 495, {Redux: A Common Interface for SOFIA Data Reduction Pipelines}, ed. A.~R. {Taylor} \& E.~{Rosolowsky}, 355

\bibitem[{{Cotera} {et~al.}(1996){Cotera}, {Erickson}, {Colgan}, {Simpson}, {Allen}, \& {Burton}}]{Cotera1996}
{Cotera}, A.~S., {Erickson}, E.~F., {Colgan}, S. W.~J., {et~al.} 1996, \apj, 461, 750, \dodoi{10.1086/177099}

\bibitem[{{Dehaes} {et~al.}(2011){Dehaes}, {Bauwens}, {Decin}, {Eriksson}, {Raskin}, {Butler}, {Dowell}, {Ali}, \& {Blommaert}}]{Dehaes2011}
{Dehaes}, S., {Bauwens}, E., {Decin}, L., {et~al.} 2011, \aap, 533, A107, \dodoi{10.1051/0004-6361/200912442}

\bibitem[{Egan {et~al.}(2003)Egan, Price, Kraemer, Mizuno, Carey, Wright, Engelke, Cohen, \& Gugliotti}]{Egan2003}
Egan, M., Price, S., Kraemer, K., {et~al.} 2003, ADA418993

\bibitem[{{Fedriani} {et~al.}(2023){Fedriani}, {Tan}, {Telkamp}, {Zhang}, {Yang}, {Liu}, {Law}, {Beltran}, {Rosero}, {Tanaka}, {Cosentino}, {Gorai}, {Farias}, {Staff}, {De Buizer}, \& {Whitney}}]{Fedriani2023}
{Fedriani}, R., {Tan}, J.~C., {Telkamp}, Z., {et~al.} 2023, \apj, 942, 7, \dodoi{10.3847/1538-4357/aca4cf}

\bibitem[{{Fritz} {et~al.}(2011){Fritz}, {Gillessen}, {Dodds-Eden}, {Lutz}, {Genzel}, {Raab}, {Ott}, {Pfuhl}, {Eisenhauer}, \& {Yusef-Zadeh}}]{Fritz2011}
{Fritz}, T.~K., {Gillessen}, S., {Dodds-Eden}, K., {et~al.} 2011, \apj, 737, 73, \dodoi{10.1088/0004-637X/737/2/73}

\bibitem[{{Ginsburg} {et~al.}(2016){Ginsburg}, {Henkel}, {Ao}, {Riquelme}, {Kauffmann}, {Pillai}, {Mills}, {Requena-Torres}, {Immer}, {Testi}, {Ott}, {Bally}, {Battersby}, {Darling}, {Aalto}, {Stanke}, {Kendrew}, {Kruijssen}, {Longmore}, \& {Dale}}]{Ginsburg2016}
{Ginsburg}, A., {Henkel}, C., {Ao}, Y., {et~al.} 2016, \aap, 586, 50, \dodoi{10.1051/0004-6361/201526100}

\bibitem[{{GRAVITY Collaboration} {et~al.}(2019){GRAVITY Collaboration}, {Abuter}, {Amorim}, {Baub{\"o}ck}, {Berger}, {Bonnet}, {Brandner}, {Cl{\'e}net}, {Coud{\'e} Du Foresto}, {de Zeeuw}, {Dexter}, {Duvert}, {Eckart}, {Eisenhauer}, {F{\"o}rster Schreiber}, {Garcia}, {Gao}, {Gendron}, {Genzel}, {Gerhard}, {Gillessen}, {Habibi}, {Haubois}, {Henning}, {Hippler}, {Horrobin}, {Jim{\'e}nez-Rosales}, {Jocou}, {Kervella}, {Lacour}, {Lapeyr{\`e}re}, {Le Bouquin}, {L{\'e}na}, {Ott}, {Paumard}, {Perraut}, {Perrin}, {Pfuhl}, {Rabien}, {Rodriguez Coira}, {Rousset}, {Scheithauer}, {Sternberg}, {Straub}, {Straubmeier}, {Sturm}, {Tacconi}, {Vincent}, {von Fellenberg}, {Waisberg}, {Widmann}, {Wieprecht}, {Wiezorrek}, {Woillez}, \& {Yazici}}]{Gravity2019}
{GRAVITY Collaboration}, {Abuter}, R., {Amorim}, A., {et~al.} 2019, \aap, 625, L10, \dodoi{10.1051/0004-6361/201935656}

\bibitem[{{Hankins} {et~al.}(2019){Hankins}, {Lau}, {Mills}, {Morris}, \& {Herter}}]{Hankins2019}
{Hankins}, M.~J., {Lau}, R.~M., {Mills}, E.~A.~C., {Morris}, M.~R., \& {Herter}, T.~L. 2019, \apj, 877, 22, \dodoi{10.3847/1538-4357/ab174e}

\bibitem[{{Hankins} {et~al.}(2020){Hankins}, {Lau}, {Radomski}, {Cotera}, {Morris}, {Mills}, {Walker}, {Barnes}, {Simpson}, {Herter}, {Longmore}, {Bally}, {Kasliwal}, {Sabha}, \& {Garc{\'\i}a-Marin}}]{Hankins2020}
{Hankins}, M.~J., {Lau}, R.~M., {Radomski}, J.~T., {et~al.} 2020, \apj, 894, 55, \dodoi{10.3847/1538-4357/ab7c5d}

\bibitem[{{Henshaw} {et~al.}(2023){Henshaw}, {Barnes}, {Battersby}, {Ginsburg}, {Sormani}, \& {Walker}}]{Henshaw2023}
{Henshaw}, J.~D., {Barnes}, A.~T., {Battersby}, C., {et~al.} 2023, in Astronomical Society of the Pacific Conference Series, Vol. 534, Protostars and Planets VII, ed. S.~{Inutsuka}, Y.~{Aikawa}, T.~{Muto}, K.~{Tomida}, \& M.~{Tamura}, 83, \dodoi{10.48550/arXiv.2203.11223}

\bibitem[{{Herter} {et~al.}(2012){Herter}, {Adams}, {De Buizer}, {Gull}, {Schoenwald}, {Henderson}, {Keller}, {Nikola}, {Stacey}, \& {Vacca}}]{Herter2012}
{Herter}, T.~L., {Adams}, J.~D., {De Buizer}, J.~M., {et~al.} 2012, \apjl, 749, L18, \dodoi{10.1088/2041-8205/749/2/L18}

\bibitem[{{Herter} {et~al.}(2013){Herter}, {Vacca}, {Adams}, {Keller}, {Schoenwald}, {Hirsch}, {Wang}, {De Buizer}, {Helton}, \& {Llorens}}]{Herter2013}
{Herter}, T.~L., {Vacca}, W.~D., {Adams}, J.~D., {et~al.} 2013, \pasp, 125, 1393, \dodoi{10.1086/674144}

\bibitem[{Hinz {et~al.}(2009)Hinz, Rieke, Yusef-Zadeh, Hewitt, Balog, \& Block}]{hinz2009}
Hinz, J., Rieke, G., Yusef-Zadeh, F., {et~al.} 2009, The Astrophysical Journal Supplement Series, 181, 227

\bibitem[{{Immer} {et~al.}(2012){Immer}, {Menten}, {Schuller}, \& {Lis}}]{Immer2012DustRidge}
{Immer}, K., {Menten}, K.~M., {Schuller}, F., \& {Lis}, D.~C. 2012, \aap, 548, A120, \dodoi{10.1051/0004-6361/201219182}

\bibitem[{{Joye} \& {Mandel}(2003)}]{JoyeMandel2003}
{Joye}, W.~A., \& {Mandel}, E. 2003, in Astronomical Society of the Pacific Conference Series, Vol. 295, Astronomical Data Analysis Software and Systems XII, ed. H.~E. {Payne}, R.~I. {Jedrzejewski}, \& R.~N. {Hook}, 489

\bibitem[{{Krabbe} {et~al.}(1991){Krabbe}, {Genzel}, {Drapatz}, \& {Rotaciuc}}]{Krabbe1991}
{Krabbe}, A., {Genzel}, R., {Drapatz}, S., \& {Rotaciuc}, V. 1991, \apjl, 382, L19, \dodoi{10.1086/186204}

\bibitem[{{Krabbe} {et~al.}(1995){Krabbe}, {Genzel}, {Eckart}, {Najarro}, {Lutz}, {Cameron}, {Kroker}, {Tacconi-Garman}, {Thatte}, {Weitzel}, {Drapatz}, {Geballe}, {Sternberg}, \& {Kudritzki}}]{Krabbe1995}
{Krabbe}, A., {Genzel}, R., {Eckart}, A., {et~al.} 1995, \apjl, 447, L95, \dodoi{10.1086/309579}

\bibitem[{{Krist} {et~al.}(2011){Krist}, {Hook}, \& {Stoehr}}]{Krist2011}
{Krist}, J.~E., {Hook}, R.~N., \& {Stoehr}, R. 2011, Proc. SPIE, 8127, 81270J, \dodoi{https://doi.org/10.1117/12.892762}

\bibitem[{{Kruijssen} \& {Longmore}(2013)}]{Kruijssen2013}
{Kruijssen}, J.~M.~D., \& {Longmore}, S.~N. 2013, \mnras, 435, 2598, \dodoi{10.1093/mnras/stt1634}

\bibitem[{{Kruijssen} {et~al.}(2014){Kruijssen}, {Longmore}, {Elmegreen}, {Murray}, {Bally}, {Testi}, \& {Kennicutt}}]{Kruijssen2014}
{Kruijssen}, J.~M.~D., {Longmore}, S.~N., {Elmegreen}, B.~G., {et~al.} 2014, \mnras, 440, 3370, \dodoi{10.1093/mnras/stu494}

\bibitem[{{Lawrence} {et~al.}(2007){Lawrence}, {Warren}, {Almaini}, {Edge}, {Hambly}, {Jameson}, {Lucas}, {Casali}, {Adamson}, {Dye}, {Emerson}, {Foucaud}, {Hewett}, {Hirst}, {Hodgkin}, {Irwin}, {Lodieu}, {McMahon}, {Simpson}, {Smail}, {Mortlock}, \& {Folger}}]{UKIDSS}
{Lawrence}, A., {Warren}, S.~J., {Almaini}, O., {et~al.} 2007, \mnras, 379, 1599, \dodoi{10.1111/j.1365-2966.2007.12040.x}

\bibitem[{{Lis} {et~al.}(1994){Lis}, {Menten}, {Serabyn}, \& {Zylka}}]{Lis1994}
{Lis}, D.~C., {Menten}, K.~M., {Serabyn}, E., \& {Zylka}, R. 1994, \apjl, 423, L39, \dodoi{10.1086/187230}

\bibitem[{{Mills} {et~al.}(2018){Mills}, {Ginsburg}, {Immer}, {Barnes}, {Wiesenfeld}, {Faure}, {Morris}, \& {Requena-Torres}}]{Mills2018}
{Mills}, E.~A.~C., {Ginsburg}, A., {Immer}, K., {et~al.} 2018, \apj, 868, 7, \dodoi{10.3847/1538-4357/aae581}

\bibitem[{{Molinari} {et~al.}(2010){Molinari}, {Swinyard}, {Bally}, {Barlow}, {Bernard}, {Martin}, {Moore}, {Noriega-Crespo}, {Plume}, {Testi}, {Zavagno}, {Abergel}, {Ali}, {Andr{\'e}}, {Baluteau}, {Benedettini}, {Bern{\'e}}, {Billot}, {Blommaert}, {Bontemps}, {Boulanger}, {Brand}, {Brunt}, {Burton}, {Campeggio}, {Carey}, {Caselli}, {Cesaroni}, {Cernicharo}, {Chakrabarti}, {Chrysostomou}, {Codella}, {Cohen}, {Compiegne}, {Davis}, {de Bernardis}, {de Gasperis}, {Di Francesco}, {di Giorgio}, {Elia}, {Faustini}, {Fischera}, {Fukui}, {Fuller}, {Ganga}, {Garcia-Lario}, {Giard}, {Giardino}, {Glenn}, {Goldsmith}, {Griffin}, {Hoare}, {Huang}, {Jiang}, {Joblin}, {Joncas}, {Juvela}, {Kirk}, {Lagache}, {Li}, {Lim}, {Lord}, {Lucas}, {Maiolo}, {Marengo}, {Marshall}, {Masi}, {Massi}, {Matsuura}, {Meny}, {Minier}, {Miville-Desch{\^e}nes}, {Montier}, {Motte}, {M{\"u}ller}, {Natoli}, {Neves}, {Olmi}, {Paladini}, {Paradis}, {Pestalozzi}, {Pezzuto}, {Piacentini}, {Pomar{\`e}s}, {Popescu}, {Reach}, {Richer}, {Ristorcelli},
  {Roy}, {Royer}, {Russeil}, {Saraceno}, {Sauvage}, {Schilke}, {Schneider-Bontemps}, {Schuller}, {Schultz}, {Shepherd}, {Sibthorpe}, {Smith}, {Smith}, {Spinoglio}, {Stamatellos}, {Strafella}, {Stringfellow}, {Sturm}, {Taylor}, {Thompson}, {Tuffs}, {Umana}, {Valenziano}, {Vavrek}, {Viti}, {Waelkens}, {Ward-Thompson}, {White}, {Wyrowski}, {Yorke}, \& {Zhang}}]{Molinari2010}
{Molinari}, S., {Swinyard}, B., {Bally}, J., {et~al.} 2010, \pasp, 122, 314, \dodoi{10.1086/651314}

\bibitem[{{Morris}(2023)}]{Morris2023}
{Morris}, M.~R. 2023, Active Galactic Nuclei, ed. F.~Combes (ISTE/Wiley), \dodoi{10.48550/arXiv.2302.02431}

\bibitem[{{Nagata} {et~al.}(1990){Nagata}, {Woodward}, {Shure}, {Pipher}, \& {Okuda}}]{Nagata1990}
{Nagata}, T., {Woodward}, C.~E., {Shure}, M., {Pipher}, J.~L., \& {Okuda}, H. 1990, \apj, 351, 83, \dodoi{10.1086/168446}

\bibitem[{{Ochsenbein} {et~al.}(2000){Ochsenbein}, {Bauer}, \& {Marcout}}]{vizier}
{Ochsenbein}, F., {Bauer}, P., \& {Marcout}, J. 2000, \aaps, 143, 23, \dodoi{10.1051/aas:2000169}

\bibitem[{{Okuda} {et~al.}(1990){Okuda}, {Shibai}, {Nakagawa}, {Matsuhara}, {Kobayashi}, {Kaifu}, {Nagata}, {Gatley}, \& {Geballe}}]{Okuda1990}
{Okuda}, H., {Shibai}, H., {Nakagawa}, T., {et~al.} 1990, \apj, 351, 89, \dodoi{10.1086/168447}

\bibitem[{{Ponti} {et~al.}(2021){Ponti}, {Morris}, {Churazov}, {Heywood}, \& {Fender}}]{Ponti2021}
{Ponti}, G., {Morris}, M.~R., {Churazov}, E., {Heywood}, I., \& {Fender}, R.~P. 2021, \aap, 646, A66, \dodoi{10.1051/0004-6361/202039636}

\bibitem[{{Ponti} {et~al.}(2019){Ponti}, {Hofmann}, {Churazov}, {Morris}, {Haberl}, {Nandra}, {Terrier}, {Clavel}, \& {Goldwurm}}]{Ponti2019}
{Ponti}, G., {Hofmann}, F., {Churazov}, E., {et~al.} 2019, \nat, 567, 347, \dodoi{10.1038/s41586-019-1009-6}

\bibitem[{{Ram{\'\i}rez} {et~al.}(2008){Ram{\'\i}rez}, {Arendt}, {Sellgren}, {Stolovy}, {Cotera}, {Smith}, \& {Yusef-Zadeh}}]{Ramirez2008}
{Ram{\'\i}rez}, S.~V., {Arendt}, R.~G., {Sellgren}, K., {et~al.} 2008, \apjs, 175, \dodoi{10.1086/524015}

\bibitem[{{Simpson}(2018)}]{Simpson2018IRS}
{Simpson}, J.~P. 2018, \apj, 857, 59, \dodoi{10.3847/1538-4357/aab55b}

\bibitem[{{Smithsonian Astrophysical Observatory}(2000)}]{SAOds92000}
{Smithsonian Astrophysical Observatory}. 2000, {SAOImage DS9: A utility for displaying astronomical images in the X11 window environment}, Astrophysics Source Code Library, record ascl:0003.002.
\newblock \doeprint{0003.002}

\bibitem[{{Stetson}(1987)}]{Stetson87}
{Stetson}, P.~B. 1987, \pasp, 99, 191, \dodoi{10.1086/131977}

\bibitem[{{Su} {et~al.}(2017){Su}, {De Buizer}, {Rieke}, {Krivov}, {L{\"o}hne}, {Marengo}, {Stapelfeldt}, {Ballering}, \& {Vacca}}]{Su2017}
{Su}, K. Y.~L., {De Buizer}, J.~M., {Rieke}, G.~H., {et~al.} 2017, \aj, 153, 226, \dodoi{10.3847/1538-3881/aa696b}

\bibitem[{{Tang} {et~al.}(2021){Tang}, {Wang}, {Wilson}, {Heyer}, {Gutermuth}, {Schloerb}, {Yun}, {Bally}, {Loinard}, {Silich}, {Ch{\'a}vez}, {Haggard}, {Monta{\~n}a}, {S{\'a}nchez-Arg{\"u}elles}, {Zeballos}, {Zavala}, \& {Le{\'o}n-Tavares}}]{Tang2021}
{Tang}, Y., {Wang}, Q.~D., {Wilson}, G.~W., {et~al.} 2021, \mnras, 505, 2392, \dodoi{10.1093/mnras/stab1191}

\bibitem[{{Wenger} {et~al.}(2000){Wenger}, {Ochsenbein}, {Egret}, {Dubois}, {Bonnarel}, {Borde}, {Genova}, {Jasniewicz}, {Lalo{\"e}}, {Lesteven}, \& {Monier}}]{SIMBAD2000}
{Wenger}, M., {Ochsenbein}, F., {Egret}, D., {et~al.} 2000, \aaps, 143, 9, \dodoi{10.1051/aas:2000332}

\bibitem[{{Whitney} {et~al.}(2008){Whitney}, {Sewilo}, {Indebetouw}, {Robitaille}, {Meixner}, {Gordon}, {Meade}, {Babler}, {Harris}, {Hora}, {Bracker}, {Povich}, {Churchwell}, {Engelbracht}, {For}, {Block}, {Misselt}, {Vijh}, {Leitherer}, {Kawamura}, {Blum}, {Cohen}, {Fukui}, {Mizuno}, {Mizuno}, {Srinivasan}, {Tielens}, {Volk}, {Bernard}, {Boulanger}, {Frogel}, {Gallagher}, {Gorjian}, {Kelly}, {Latter}, {Madden}, {Kemper}, {Mould}, {Nota}, {Oey}, {Olsen}, {Onishi}, {Paladini}, {Panagia}, {Perez-Gonzalez}, {Reach}, {Shibai}, {Sato}, {Smith}, {Staveley-Smith}, {Ueta}, {Van Dyk}, {Werner}, {Wolff}, \& {Zaritsky}}]{Whitney2008}
{Whitney}, B.~A., {Sewilo}, M., {Indebetouw}, R., {et~al.} 2008, \aj, 136, 18, \dodoi{10.1088/0004-6256/136/1/18}

\bibitem[{{Young} {et~al.}(2012){Young}, {Becklin}, {Marcum}, {Roellig}, {De Buizer}, {Herter}, {G{\"u}sten}, {Dunham}, {Temi}, {Andersson}, {Backman}, {Burgdorf}, {Caroff}, {Casey}, {Davidson}, {Erickson}, {Gehrz}, {Harper}, {Harvey}, {Helton}, {Horner}, {Howard}, {Klein}, {Krabbe}, {McLean}, {Meyer}, {Miles}, {Morris}, {Reach}, {Rho}, {Richter}, {Roeser}, {Sandell}, {Sankrit}, {Savage}, {Smith}, {Shuping}, {Vacca}, {Vaillancourt}, {Wolf}, \& {Zinnecker}}]{Young2012}
{Young}, E.~T., {Becklin}, E.~E., {Marcum}, P.~M., {et~al.} 2012, \apjl, 749, L17, \dodoi{10.1088/2041-8205/749/2/L17}

\bibitem[{{Zhao} {et~al.}(2016){Zhao}, {Morris}, \& {Goss}}]{Zhao2016}
{Zhao}, J.-H., {Morris}, M.~R., \& {Goss}, W.~M. 2016, \apj, 817, 171, \dodoi{10.3847/0004-637X/817/2/171}

\end{thebibliography}
\bibliographystyle{aasjournal}

\appendix
\section{The {\it{SOFIA/FORCAST}} Galatic Center Catalogs}\label{howto}

The \sofia/FORCAST Galactic Center catalogs are available for download on \textit{IRSA} along with a searchable form via the gator database. Additional versions are available as fits tables hosted on Zenodo along with the source code developed for the project. This file is named `SFGC\_MasterCatalog.fits', and includes all the information presented in Table \ref{full_cat}. Table \ref{full_cat} provides the exact nomenclature for each column, any units associated with the column data, and a brief description of the measurements included in that column.  The fits table version of the catalog also contains cutout images for each source (such as those presented in Figures \ref{sources25}-\ref{irregular}), and residual images after subtracting the Moffat and Gaussian fits where applicable, and therefore is a rather large file. An additional file named `SFGC\_MasterCatalog\_NoCutouts.fits' contains all the same measurements, but excludes the cutout and residual images. Last, `SFGC\_ShortCatalog.fits' contains only the measurements illustrated in Table \ref{short_cat}; the column nomenclature and descriptions are provided in Table \ref{short_cat_tab}. Additional ascii versions of `SFGC\_MasterCatalog\_NoCutouts' and `SFGC\_ShortCatalog' are also provided. 

\startlongtable
\begin{deluxetable*} {l|c|l} 
\tabletypesize{\footnotesize}
\tablecaption{Master Catalog Data Nomenclature\label{full_cat}}
\tablehead{
\colhead{Column Name}&\colhead{Units}&\colhead{Description} 
}
\startdata
SourceID & &{\bf{S}}ofia {\bf{F}}ORCAST {\bf{G}}alactic {\bf{C}}enter~{\it{l}}~{\it{b}}  i.e. SFGC0.036-0.1818\\
ra	&	deg	&	Right Ascension	(J2000)\\
dec	&	deg	&	Declination	(J2000)\\
BestModel\_25um&&Best estimate of source flux: A=Aperture, M=Moffat, G=Gaussian\\
Flux\_25um\_4pix&Jy&Background subtracted 4 pixel aperture flux\\
Flux\_25um\_4pix\_err&Jy&4 pixel aperture flux uncertainty\\
SNR\_25um\_4pix& &SNR for 4 pixel aperture flux\\
Flux\_25um\_8pix&Jy&Background subtracted 8 pixel aperture flux\\
Flux\_25um\_8pix\_err&Jy&8 pixel aperture flux uncertainty\\
SNR\_25um\_8pix& &SNR for 8 pixel aperture flux\\
Flux\_25um\_12pix&Jy&Background subtracted 12 pixel aperture flux\\
Flux\_25um\_12pix\_err&Jy&12 pixel aperture flux uncertainty\\
SNR\_25um\_12pix& &SNR for 12 pixel aperture flux\\
ApRatio\_25um\_8\_4&&Ratio of the 8 pixel to 4 pixel aperture fluxes\\
ApRatio\_25um\_12\_8&&Ratio of the 12 pixel to 8 pixel aperture fluxes\\
ApPhot\_25um\_qflag&\#\#\#&Quality flags for 25 \micron\ photometry:  $0=$~SNR~$>3$,  $1=$~SNR~$<3$  \\
&& 4 pixel aperture/8 pixel aperture/12 pixel aperture \\
UL\_25um\_4pix&Jy&4 pixel aperture measured upper limit for source identified in 37 \micron\ observations\\
UL\_25um\_8pix&Jy&8 pixel aperture measured upper limit for source identified in 37 \micron\ observations\\
UL\_25um\_12pix&Jy&12 pixel aperture measured upper limit for source identified in 37 \micron\ observations\\
Flux\_25um\_M2D&Jy&Measured 2D Moffat fit flux\\
Flux\_25um\_M2D\_err&Jy&2D Moffat fit flux uncertainty\\
SNR\_25um\_M2D& &2D Moffat fit signal-to-noise measurement\\
chi2\_25um\_M2D&&$\chi^2$ measurement of  2D Moffat fit\\ 
qflag\_25um\_M2D&&Quality flag for the 2D Moffat fitting procedure. 0 = Nominal, 1 = Failed \\
Flux\_25um\_G2D&Jy&Measured 2D Gaussian fit flux\\
Flux\_25um\_G2D\_err&Jy&2D Gaussian fit flux uncertainty\\
SNR\_25um\_G2D& &2D Gaussian fit signal-to-noise measurement\\
chi2\_25um\_G2D&&$\chi^2$ measurement of  2D Gaussian fit\\
qflag\_25um\_G2D&&Quality flag for the 2D Gaussian fitting procedure. 0 = Nominal, 1 = Failed  \\
fwhm\_25um&pixel&2D Gaussian fit derived FWHM\\
fwhm\_25um\_err&pixel&2D Gaussian FWHM measured uncertainty\\
elong\_25um&&2D Gaussian fit derived elongation parameter (ratio of semimajor and semiminor axes)\\
vExtFlag\_25um && Very extended flag at 25 \micron. True if measured FWHM$>$12 pix, False if $\leq$12 pix \\
bkg\_qflag\_25um&&Quality flag for background measurement. 0 = nominal, 1 = high local background \\
edge\_flag\_25um&& True if the source is located in a dithered region with $<$80\% of the total exposure time, False if nominal \\
Err\_CalF\_25um&& 25 \micron~ calibration error factor, $\eta_{flux}$. See section \ref{uncertainties} \\
cutout\_25um&pix,pix&Central coordinates for 25\arcsec$\times$25\arcsec cutout image at 25 \micron \\
M2D\_resid\_25um&pix,pix&Residual image produced by subtracting the Moffat model from the data cutout\\
G2D\_resid\_25um&pix,pix&Residual image produced by subtracting the Gaussian model from the data cutout \\ 
BestModel\_37um&&Best estimate of source flux: A=Aperture, M=Moffat, G=Gaussian \\
Flux\_37um\_5\_5pix&Jy&Background subtracted 5.5 pixel aperture flux\\
Flux\_37um\_5\_5pix\_err&Jy&5.5 pixel aperture flux uncertainty\\
SNR\_37um\_5\_5pix& &SNR for 5.5 pixel aperture flux\\
Flux\_37um\_10\_5pix&Jy&Background subtracted 10.5 pixel aperture flux\\
Flux\_37um\_10\_5pix\_err&Jy&10.5 pixel aperture flux uncertainty\\
SNR\_37um\_10\_5pix& &SNR for 10.5 pixel aperture flux\\
Flux\_37um\_14pix&Jy&Background subtracted 14 pixel aperture flux\\
Flux\_37um\_14pix\_err&Jy&14 pixel aperture flux uncertainty\\
SNR\_37um\_14pix& &SNR for 14 pixel aperture flux\\
ApRatio\_37um\_10\_5&&Ratio of the 10.5 pixel to 5.5 pixel aperture fluxes\\
ApRatio\_37um\_14\_10&&Ratio of the 14 pixel to 10.5 pixel aperture fluxes\\
ApPhot\_37um\_qflag&\#\#\#&Quality flags for 37 \micron\ photometry:  $0=$~SNR~$>3$,  $1=$~SNR~$<3$  \\
&& 5.5 pixel aperture/10.5 pixel aperture/14 pixel aperture \\
UL\_37um\_5\_5pix&Jy&5.5 pixel aperture measured upper limit for source identified in 25 \micron\ observations\\
UL\_37um\_10\_5pix&Jy&10.5 pixel aperture measured upper limit for source identified in 25 \micron\ observations\\
UL\_37um\_14pix&Jy&14 pixel aperture measured upper limit for source identified in 25 \micron\ observations\\Flux\_37um\_M2D&Jy&Measured 2D Moffat fit flux\\
Flux\_37um\_M2D\_err&Jy&2D Moffat fit flux uncertainty\\
SNR\_37um\_M2D& &2D Moffat fit signal-to-noise measurement\\
chi2\_37um\_M2D&&$\chi^2$ measurement of  2D Moffat fit\\ 
qflag\_37um\_M2D&&Quality flag for the 2D Moffat fitting procedure. 0 = Nominal, 1 = Failed \\
Flux\_37um\_G2D&Jy&Measured 2D Gaussian fit flux\\
Flux\_37um\_G2D\_err&Jy&2D Gaussian fit flux uncertainty\\
SNR\_37um\_G2D& &2D Gaussian fit signal-to-noise measurement\\
chi2\_37um\_G2D&&$\chi^2$ measurement of 2D Gaussian fit\\
qflag\_37um\_G2D&&Quality flag for the 2D Gaussian fitting procedure. 0 = Nominal, 1 = Failed \\
fwhm\_37um&pixel&2D Gaussian fit derived FWHM\\
fwhm\_37um\_err&pixel&2D Gaussian fit measured uncertainty\\
elong\_37um&&2D Gaussian fit derived elongation parameter (ratio of semimajor and semiminor axes)\\
vExtFlag\_37um &&Very extended flag at 37 \micron. True if measured FWHM$>$14 pix, False if $\leq$14 pix  \\
bkg\_qflag\_37um&&Quality flag for background measurement. 0 = nominal, 1 = high local background \\
edge\_flag\_37um&&True if the source is located in a dithered region with $<$80\% of the total exposure time, False if nominal\\
Err\_CalF\_37um&& 37 \micron~ calibration error factor, $\eta_{flux}$. See section \ref{uncertainties} \\
cutout\_37um&pix,pix&Central coordinates for 30\arcsec$\times$30\arcsec cutout image at 37 \micron \\
M2D\_resid\_37um&pix,pix&Residual image produced by subtracting the Gaussian model from the data cutout \\
G2D\_resid\_37um&pix,pix&Residual image produced by subtracting the Gaussian model from the data cutout \\ 
FieldID&&Observational Field used to measure source, corresponds to Figure 1 and Table 1 nomenclature\\
SID25&&Source ID name based on measured  25 \micron\ coordinates\\
SID37&&Source ID name based on measured  37 \micron\ coordinates\\
Found25 && True if source initially detected only at 25 \micron  \\
Found37 && True if source initially detected only at 37 \micron \\
Matched && True if source detected at both 25 and 37 \micron \\
F25\_F37\_sep & arcsec & Separation distance between 25 and 37 \micron\ sources that are considered `Matched' \\
SSTGCxmatch&&SSTGC catalog source number matched to 25 \micron\ data \\
SSTGCd2d & arcsec & Separation distance between the FORCAST source coordinates and the matched SSTGC catalog source \\
tau\_9\_6 & &Estimated $\tau_{9.6~\micron}$ extinction value \\
tau\_distance & arcsec & Distance to nearest observed $\tau_{9.6~\micron}$ source \\
\hline
\enddata
\end{deluxetable*}

\startlongtable
\begin{deluxetable*} {l|c|l} 
\tabletypesize{\footnotesize}
\tablecaption{Short Catalog Data Nomenclature\label{short_cat_tab}}
\tablehead{
\colhead{Column Name}&\colhead{Units}&\colhead{Description} 
}
\startdata
SourceID & &{\bf{S}}ofia {\bf{F}}ORCAST {\bf{G}}alactic {\bf{C}}enter~{\it{l}}~{\it{b}}  i.e. SFGC0.036-0.1818\\
RA(J2000)	&	hms	&	Right Ascension	\\
DEC(J2000)	&	dms	&	Declination	\\
Flux\_25um\_8pix&Jy&Background subtracted 8 pixel aperture flux\\
Flux\_25um\_8pix\_err&Jy&8 pixel aperture flux uncertainty\\
Flux\_25um\_G2D&Jy&Measured 2D Gaussian fit flux\\
Flux\_25um\_G2D\_err&Jy&2D Gaussian fit flux uncertainty\\
fwhm\_25um&pixel&2D Gaussian fit derived FWHM\\
Flux\_37um\_10\_5pix&Jy&Background subtracted 10.5 pixel aperture flux\\
Flux\_37um\_10\_5pix\_err&Jy&10.5 pixel aperture flux uncertainty\\
Flux\_37um\_G2D&Jy&Measured 2D Gaussian fit flux\\
Flux\_37um\_G2D\_err&Jy&2D Gaussian fit flux uncertainty\\
fwhm\_37um&pixel&2D Gaussian fit derived FWHM\\
\hline
\enddata
\end{deluxetable*}

\section{Code Developed for Creation of the {\it{SOFIA/FORCAST}} Galatic Center Catalog}\label{code}

Code developed for creating the SFGC catalog is available on Zenodo\footnote{\url{https://doi.org/10.5281/zenodo.11459088}} as well as the project page on GitHub\footnote{\url{https://github.com/mjhankins/SFGCphotcode}}. The code is based largely on the Astropy Project suite of Python packages \citep{astropy2013,astropy2018,astropy22} and Photutils \citep{larry_bradley_2020_4044744}. As part of the SOFIA Galactic Center Legacy Program, this code was developed to provide a general photometry tool for FORCAST data, and this appendix provides information using the code. The software links contain a series of notebooks and scripts that were used to produce the catalog. These items are briefly described below, with additional information provided online with the above mentioned software repositories:
\begin{itemize}
    \item Config.py - A configuration file that defines properties for FORCAST filters and data files to be used by the code. Definitions for photometry routines, such as aperture radii are set in this file. A user can also specify the file path to SAODds9 on their machine to take advantage of interactive features in the notebooks and scripts. Note that in the online version of the code, this file is named `config\_example.py' and must be renamed to `config.py' with appropriate adjustments to the file contents for it to work on a particular machine. 
    
    \item SourceCatalog\_Detect.ipynb - This notebook contains the routines for source detection used in the catalog. The notebook is able to process one observation at a time, and the user can select which file to use from those listed in the config.py file. The code processes the data file following the same steps described in the main text - First 2D background subtraction is performed to eliminate large scale structure on the images. Next, DAOstarfinder routine from photutils is then used to identify sources, followed by the Segmentation map routine. After these routines run, there is an option to include user defined sources, which can be marked as regions in an interactive ds9 window. Any saved regions regions file with the field name from the config file appended with '\_ds9.reg' will be loaded in the code as user define source. From here, there is a step that combines the results from the DAOstarfinder routine, Segmentation map, and user sources into a singular source list. For sources found via multiple methods, the priority for keeping a source is as followed: User, DAO, Segmentation. The final product of this notebook is a fits file named indicating field name, wavelength, and ending in `\_CombinedSources.fits'
    
    The online repository also includes a script version of this notebook named `SourceCatalog\_Detect.py' which will run a batch job of the steps in the notebook for every file defined in the `config.py' file for a specified filter.

    \item SourceCatalog\_ApPhot.ipynb - This notebook contains the routines for source photometry (both aperture and model) used in the catalog. This notebook is able to process one observation at a time, and the user can select which file to use from those listed in the config.py file. The code processes the data file following the same steps described in the main text - 2D background subtraction is performed to eliminate large scale structure on the images. Aperture photometry is then performed on all sources in the fits file produced by the source detection notebook/script, and an annulus aperture is used to measure local background for each source. Next, background subtracted photometry is produced at each radii defined in the cofig.py file. After that the source list is passed to the routines for fitting the Moffat and Gaussian profiles. Both the Moffat and Gaussian fit is attempted for every source, but if the fit fails, the parameters for that fit is returned as `NaN'. Additional metrics are performed based on the covariance matrix produced by the fit and also reduced $\chi^2$ which aid in determining if a particular source model is better than the other, or if neither produces a satisfactory fit. This characterization is indicated in the subsequently produced table as the 'BestModel' flag. 

    Similar to the detection notebook, the photometry notebook also comes with a script form named `SourceCatalog\_ApPhot.py' which will run a batch job of the photometry at a given wavelength for every file defined in the 'config.py' file.
    
    \item SourceCatalog\_CombineFields.ipynb - This notebook combines the outputs of the aperture photometry notebook/script to a singular table for each filter defined in the config.py file. The script checks for duplication of sources which might be caused by overlapping coverage between fields. The default cross-match radius for source removal is 3\arcsec. In the even duplication is detected, the higher SNR source is kept. After this point, there is also option for user removed sources, which may occur for duplication outside of the 3\arcsec\ match radius, caused by source near the edge of fields where astrometry is less precise.  Next, the SNR quality cut is applied, and there is an option to remove visually bad sources which may pass the SNR cut but be related to detector artifacts or other types of spurious sources. This process is repeated for both the 25 and 37 micron filter, and then the source tables at both wavelengths are cross-matched to create a singular table of sources at both wavelengths. Any sources that are not detected at both wavelengths are flagged and forced photometry is run at the non-detected wavelength. Any sources at this step that do not meet the SNR$>$3 threshold are flagged, given NULL values and 3-$\sigma$ upper limits are calculated in a separate column. The final source table produce by this script is a fits file ending in  `\_step3.fits'.

    \item SourceCatalog\_SSTGC\_Xmatch.ipynb - This notebook reads in the `\_step3.fits' file from earlier and performs a cross-match based on source coordinates to the SSTGC catalog. This notebook can be easily modified to reference other source catalogs present in VizieR, which gives a broader use case for this notebook. The script cross-match is performed at a 3\arcsec\ radius, which was selected as the 2-$\sigma$ astrometric uncertainty for the catalog (See Figure 7). This script can also be used to produce Figure 7 in order to study the asymmetric uncertainty of a particular FORCAST data set compared to another catalog. The final product produced by this notebook is a file `\_step4.fits', which is the final product of the catalog. Two versions of the catalog are produced, one which includes all source image cutouts and one that does not. The two version approach is meant to help with file size as the cutouts take up a large amount of space and may not be needed in every use case. Though the cutouts are likely useful for assessing quality of sources in most cases. 

    \item ForcastPhot.py - This file contains the functions referenced in the above notebooks/scripts including functions for performing background subtracted aperture photometry, routines for fitting 2D Gaussian models and Moffat Models to source cutouts, functions for combining source lists between data sets, removing duplicated sources found within a defined radius, and creating ds9 region files to aid in examining sources included in the catalog. 
\end{itemize}

\end{document}